\documentclass[twocolumn,  tighten,  twocolappendix]{aastex631}

\shorttitle{Rotation Curves of SDSS-IV MaNGA Galaxies}
\shortauthors{Yoon et al. (2021)}

\begin{document}

\title{Rotation Curves of Galaxies and Their Dependence on Morphology and Stellar Mass}

%\correspondingauthor{Haeun Chung (haeunchung@arizona.edu)}
\email{yyoon@kias.re.kr}

\author[0000-0003-0134-8968]{Yongmin Yoon}
\affiliation{School of Physics, Korea Institute for Advanced Study (KIAS), 85, Hoegiro, Dongdaemun-gu, Seoul, 02455, Republic of Korea}

\author[0000-0001-9521-6397]{Changbom Park}
\affiliation{School of Physics, Korea Institute for Advanced Study (KIAS), 85, Hoegiro, Dongdaemun-gu, Seoul, 02455, Republic of Korea}

\author[0000-0002-3043-2555]{Haeun Chung}
\affiliation{University of Arizona, Steward Observatory, 933 N Cherry Ave, Tucson, AZ 85721, USA}

\author[0000-0002-9808-3646]{Kai Zhang}
\affiliation{Lawrence Berkeley National Laboratory, 1 Cyclotron Rd, Berkeley, CA 94720, USA}

\begin{abstract}
We study how stellar rotation curves (RCs) of galaxies are correlated on average with morphology and stellar mass ($M_\mathrm{star}$) using the final release of Sloan Digital Sky Survey IV MaNGA data. We use the visually assigned $T$-types for the morphology indicator, and adopt a functional form for the RC that can model non-flat RCs at large radii. We discover that within the radial coverage of the MaNGA data, the popularly known flat rotation curve at large radii applies only to the particular classes of galaxies, i.e., massive late types ($T$-type $\geq1$, $M_\mathrm{star}\gtrsim10^{10.8}M_\odot$) and S0 types ($T$-type$=-1$ or $0$, $M_\mathrm{star}\gtrsim10^{10.0}M_\odot$). The RC of late-type galaxies at large radii rises more steeply as $M_\mathrm{star}$ decreases, and its slope increases to about $+9$ km s$^{-1}$kpc$^{-1}$ at $M_\mathrm{star}\approx10^{9.7}M_\odot$. By contrast, elliptical galaxies ($T$-type $\le-2$) have descending RCs at large radii. Their slope becomes more negative as $M_\mathrm{star}$ decreases, and reaches as negative as $-15$ km s$^{-1}$kpc$^{-1}$ at $M_\mathrm{star}\approx10^{10.2}M_\odot$. We also find that the inner slope of the RC is highest for elliptical galaxies with $M_\mathrm{star}\approx10^{10.5}M_\odot$, and decreases as $T$-type increases or $M_\mathrm{star}$ changes away from $10^{10.5}M_\odot$. The velocity at the turnover radius $R_t$ is higher for higher $M_\mathrm{star}$, and $R_t$ is larger for higher $M_\mathrm{star}$ and later $T$-types. We show that the inner slope of the RC is coupled with the central surface stellar mass density, which implies that the gravitational potential of central regions of galaxies is dominated by baryonic matter. With the aid of simple models for matter distribution, we discuss what determines the shapes of RCs.
\end{abstract}


\section{Introduction}\label{sec:intro}

The rotation curve (RC) of a galaxy is the rotating speed of stars or gas in the galaxy as a function of radial distance from the galaxy center. The exploration to investigate galaxy RCs began with nearby galaxies such as M31. \citet{Babcock1939} studied the RC of M31 and discovered nearly constant rotation velocities at the outer part of the galaxy. \citet{Babcock1939} suggested that the mass-to-light ratio should increase at the outer part of the galaxy to explain the RC. \citet{Oort1940} also arrived at a similar conclusion based on NGC 3115, suggesting that substantial invisible matter\footnote{The invisible matter was presumed to be extremely faint dwarf stars or interstellar dust and gas in \citet{Oort1940}.} should be in the outer part of the galaxy. Radio observations for \ion{H}{1} gas also detected flat or nearly flat RCs in M31 \citep{vandeHulst1957} and M33 \citep{Volders1959}. 

With the improvement of observational instruments, it was found that flat RCs, which cannot be fully explained by visible or detectable matters, were common properties in many galaxies \citep{Rubin1980,Persic1996,Catinella2006}. This implies that Newtonian dynamics needs to be modified on the galactic scale \citep{Milgrom1983a,Milgrom1983b} or invisible matter (dark matter) that we have not identified yet is prevalent in the universe \citep{Trimble1987,deSwart2017}. For now, dark matter is an indispensable factor that describes the current standard cosmology. In this way, RCs have been used as a tool for determining mass distributions in galaxies. 

Many previous studies investigated the relation between the shape of RCs and various galaxy properties in order to understand the correlation between dynamical mass distributions in galaxies and their properties such as morphology, structure, and luminosity.  For example, the shape of the RC is coupled with the galaxy luminosity in the sense that rising outer RC profiles are commonly found in low-luminosity galaxies, while high-luminosity galaxies show flat or even declining RCs \citep{Casertano1991,Persic1996,Sofue2001,Noordermeer2007,Swaters2009,Kalinova2017}. Some studies showed that the shape of the RC is dependent on galaxy morphology \citep{Corradi1990,Casertano1991,Noordermeer2007,Swaters2009,Erroz2016,Kalinova2017}. Combining and summarizing these results, early-type galaxies generally show steeply rising inner and declining/flat outer RC profiles, while late-type galaxies have relatively slowly rising inner and flat/rising outer RC profiles. Some studies could not find such a trend for RCs, depending on morphology \citep{Burstein1985,Rubin1985}.

The advent of integral field unit (IFU) spectroscopy allows us to obtain spatially resolved spectra in a specific region of the sky with relatively low observation costs. Thus, observational surveys using IFU spectroscopy are capable of providing a lot of spatially resolved kinematic information for numerous galaxies.  Many IFU surveys have been conducted in the recent decade, such as the SAURON project \citep{Bacon2001}, the MASSIVE survey \citep{Ma2014}, the DiskMass survey \citep{Bershady2010}, the ATLAS$^\mathrm{3D}$ project \citep{Cappellari2011}, the CALIFA survey \citep{Sanchez2012,Sanchez2016}, and the SAMI survey \citep{Croom2021}. The numbers of observed galaxies in these surveys range from tens to hundreds, except the SAMI survey that observed $\sim3000$ galaxies.

The MaNGA (Mapping Nearby Galaxies at Apache Point Observatory) IFU survey \citep{Bundy2015,Drory2015,Yan2016,Wake2017} is a project of the fourth generation of the Sloan Digital Sky Survey (SDSS; \citealt{Blanton2017}) using the ARC 2.5 m telescope \citep{Gunn2006}. More than $\sim10,000$ galaxies were observed in the MaNGA project. This unprecedented number of observed galaxies makes it possible to conduct statistical studies of spatially resolved kinematics for a large amount of diverse galaxies that were uniformly observed.    

Previous studies on the relation between the shape of RCs and galaxy properties are not based on as large a number of uniformly observed galaxies as the MaNGA data have. Now, a comprehensive understanding of shapes of RCs across various types of galaxies is possible with the aid of the MaNGA data. Here, we investigate the relation between shapes of stellar RCs for MaNGA galaxies and galaxy properties such as stellar mass ($M_\mathrm{star}$) and morphology, and we discuss how the dark matter and baryon mass distributions in galaxies are coupled with morphology and $M_\mathrm{star}$ of galaxies.

 Throughout this study, we use \emph{H$_0=70$} km s$^{-1}$ Mpc$^{-1}$, $\Omega_{\Lambda}=0.7$, and $\Omega_\mathrm{m}=0.3$ as cosmological parameters. 
\\

\section{Sample and Analysis}\label{sec:sample}
\subsection{SDSS-IV MaNGA}\label{sec:manga}
The MaNGA project uses 17 fiber-bundled IFUs with sizes of $12\arcsec$ to $32\arcsec$ depending on the number of fibers. These 17 IFUs are assigned within the $3\degr$ field of view. The spectrograph of the MaNGA survey is the same as that of the Baryon Oscillation Spectroscopic Survey \citep{Smee2013}. It covers a wavelength range of 3,600--10,300\AA\, with a mid-range spectral resolution of $R\sim2000$. The target galaxy samples of MaNGA were selected by using criteria for $i$-band absolute magnitude and redshift (and near-ultraviolet -- $i$ color for a small portion of targets) that result in $\sim10,000$ galaxies evenly distributed in the color-magnitude space and having an uniform spectroscopic coverage up to 1.5 or 2.5 half-light radius along the major axis ($R_e$). See \citet{Wake2017} for details about selecting target galaxy samples.

In this study, we use IFU data cubes from the MaNGA internal data release MPL-11, which is equivalent to the final release version of the MaNGA data, consisting of all observational data obtained during the survey. There are 10,010 galaxies with high-quality observations in this final release version. They are in the redshift range of $z<0.15$ with a median value of $z=0.037$ and in the $M_\mathrm{star}$ range of $M_\mathrm{star}\gtrsim 10^9 M_\odot$ with a median value of  $M_\mathrm{star}=10^{10.5} M_\odot$. Only $\sim1\%$ of MaNGA data were repeated observations for the same galaxy. We used one observation among these duplicated ones by choosing the data with the largest IFU size. If the IFU sizes are identical, then we chose the data with the highest blue channel signal-to-noise ratio (S/N). We note that $M_\mathrm{star}$ and $R_e$ of galaxies used here are based on the 2D S{\'e}rsic model in the NASA-Sloan Atlas (NSA) catalog\footnote{This is a base catalog for the selection of target galaxies in the MaNGA survey \citep{Wake2017}.} ($R_e$ was measured in the $r$ band). 
\\

\subsection{Deconvolution of IFU Data}\label{sec:decon}
Data from ground-based telescopes are bound to be largely affected by atmospheric seeing as well as aberration from the instruments. The seeing effect is more severe in the data from fiber-based IFUs due to large physical gaps between sampling elements (IFU fibers). Minimizing such an effect allow us to extract more robust spatially resolved kinematics, particularly in the central parts of galaxies. As in \citet{Chung2020}, we deconvolved all of the IFU data cubes of 10,010 galaxies using the Lucy--Richardson (LR) algorithm \citep{Richardson1972,Lucy1974} in order to mitigate the seeing effect. The LR deconvolution algorithm is an iterative procedure to recover an original image that has been convolved by a point spread function (PSF). This algorithm requires only a few parameters to perform, so that it is suitable for deconvolution of very large data. Details about the algorithm and its application to MaNGA data cubes are in \citet{Chung2020}. Therefore, in this paper, we only briefly describe the implementation of the LR algorithm on the IFU data. 

The algorithm can be described by a simple equation \citep{Shepp1982,Chung2020},
\begin{equation}
u^{n+1}=u^n\cdot \bigg(\frac{d}{u^n\otimes p}\otimes p\bigg),
\label{eq:deconv}
\end{equation}
where $u^n$ is $n\mathrm{th}$ estimation of the maximum likelihood solution and $d$ is an original PSF-convolved image (hence $u^0=d$). Parameter $p$ is a two-dimensional (2D) PSF and $\otimes$ means 2D convolution. This LR algorithm was applied to the MaNGA cube data in a sense that the 2D image slice at each wavelength bin was deconvolved separately. MaNGA data cubes contain PSF full-width-half-maximum (FWHM) values at $g$, $r$, $i$, and $z$ bands. We fit a linear function to these FWHM values at the wavelengths of the four bands. The FWHM of the PSF at each wavelength bin in the data cubes was determined from this linear function,\footnote{The difference between FWHM values of the four bands recorded in MaNGA data cubes and those from the fitted linear function is negligible: the average absolute difference is $0.007\arcsec$ \citep{Chung2020}, since the FWHM value in MaNGA data is only a weak function of wavelength \citep{Law2016,Chung2020}. Thus, the FWHM value derived from the linear function is a good approximation at each wavelength bin.}  and a 2D Gaussian function with that FWHM was used in the deconvolution process.\footnote{We note that PSFs of MaNGA data are well described by a single 2D Gaussian function and the FWHM varies less than $10\%$ across a given IFU \citep{Law2015,Law2016}.} The number of iterations ($N_\mathrm{iter}$) in the LR algorithm was set to 20.  This is because $N_\mathrm{iter}=20$ is the optimum iteration number as tested in \citet{Chung2020}: beyond $N_\mathrm{iter}=20$, the quality of deconvolution is not significantly improved further, while additional artifacts can occur in the image with amplified noise.  

\citet{Chung2020} intensively tested the LR deconvolution algorithm using MaNGA IFU data and mock IFU data. Specifically, \citet{Chung2020} generated a great deal of mock data of which the properties are identical to MaNGA IFU data in terms of the field of view, radial coverage, S/N, PSF effects, etc. To mimic realistic cases, galaxies with various surface brightness profiles (S\'{e}rsic indices) and diverse RC model parameters (Equation \ref{eq:rc1}) were assigned to the mock IFU data. Then, they deconvolved the mock IFU data and derived RC parameters from the data. By comparing the derived RC parameters and the true values, they demonstrated that the LR algorithm effectively recovers the true stellar kinematics of galaxies. For example, the inner slope of the RC (Equation \ref{eq:sin}), which is expected to be largely influenced by PSF effects, can be recovered with an underestimation less than $\sim25\%$. We note that the underestimation of the inner slope of the RC due to the limitation of the deconvolution process varies less than $\sim15\%$ between galaxies in different $M_\mathrm{star}$ bins.
\\

\begin{figure*}
\includegraphics[width=\linewidth]{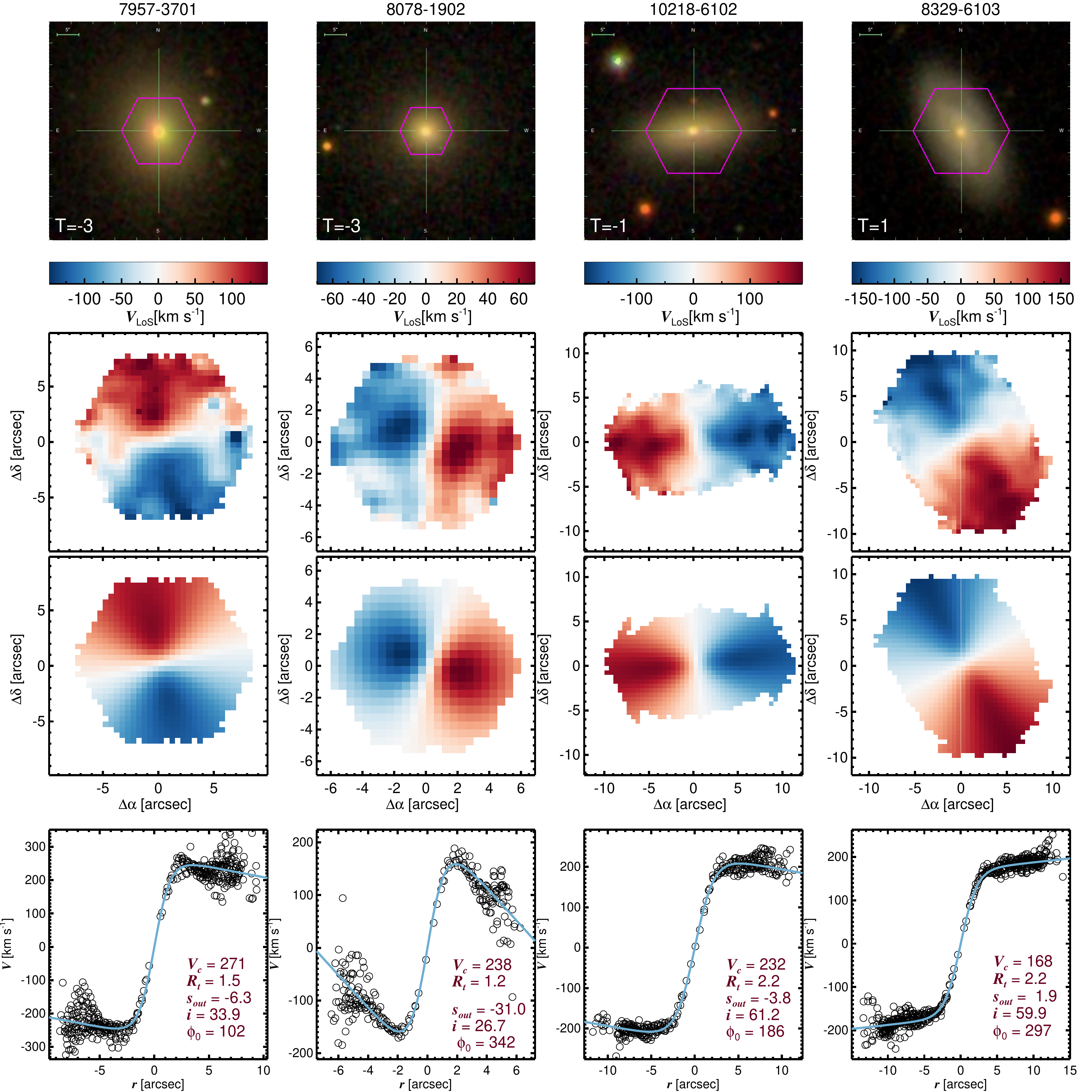}  %width=\linewidth
\centering
\caption{Examples of galaxies used in this study. The first row: color images in SDSS. The Plate ID (e.g., 7957) and IFU design ID (e.g., 3701) are on top of the color images. The hexagon represents the MaNGA IFU field of view. $T$-type values of the galaxies are in the lower left corner of the images. The second row: extracted 2D line-of-sight velocity maps in which systematic line-of-sight velocities at the kinematic center are set to 0 km s$^{-1}$. See the color bars on top of the velocity maps for the color-coded line-of-sight velocity scales. The $x$ and $y$ axes are relative R.A. and decl., respectively. The third row: the best-fit 2D models (Equation \ref{eq:rc2}) for the observed velocity maps. The bottom rows: observed 1D velocity profiles. The circles are line-of-sight velocities at $r$ (distance from the kinematic center) after correcting for the geometric term $\cos(\phi-\phi_0)$ and the kinematic inclination $\sin i$ (shown here are data points derived from spectra with median S/N$\ge10$ and located within $\pm60\degr$ from the kinematic major axis). The solid lines are the best-fit rotation curve models of Equation \ref{eq:rc1}. In each panel of the bottom rows, we provide the fitted model parameters of Equations \ref{eq:rc1} and \ref{eq:rc2} ($V_c$ in km s$^{-1}$, $R_t$ in arcseconds, $s_\mathrm{out}$ in km s$^{-1}$arcsec$^{-1}$, and $i$, $\phi_0$ in degrees). These examples are galaxies with $T$-type $\le1$. The corresponding section: Section \ref{sec:sample}.
\label{fig:ex1}}
\end{figure*}

\begin{figure*}
\includegraphics[width=\linewidth]{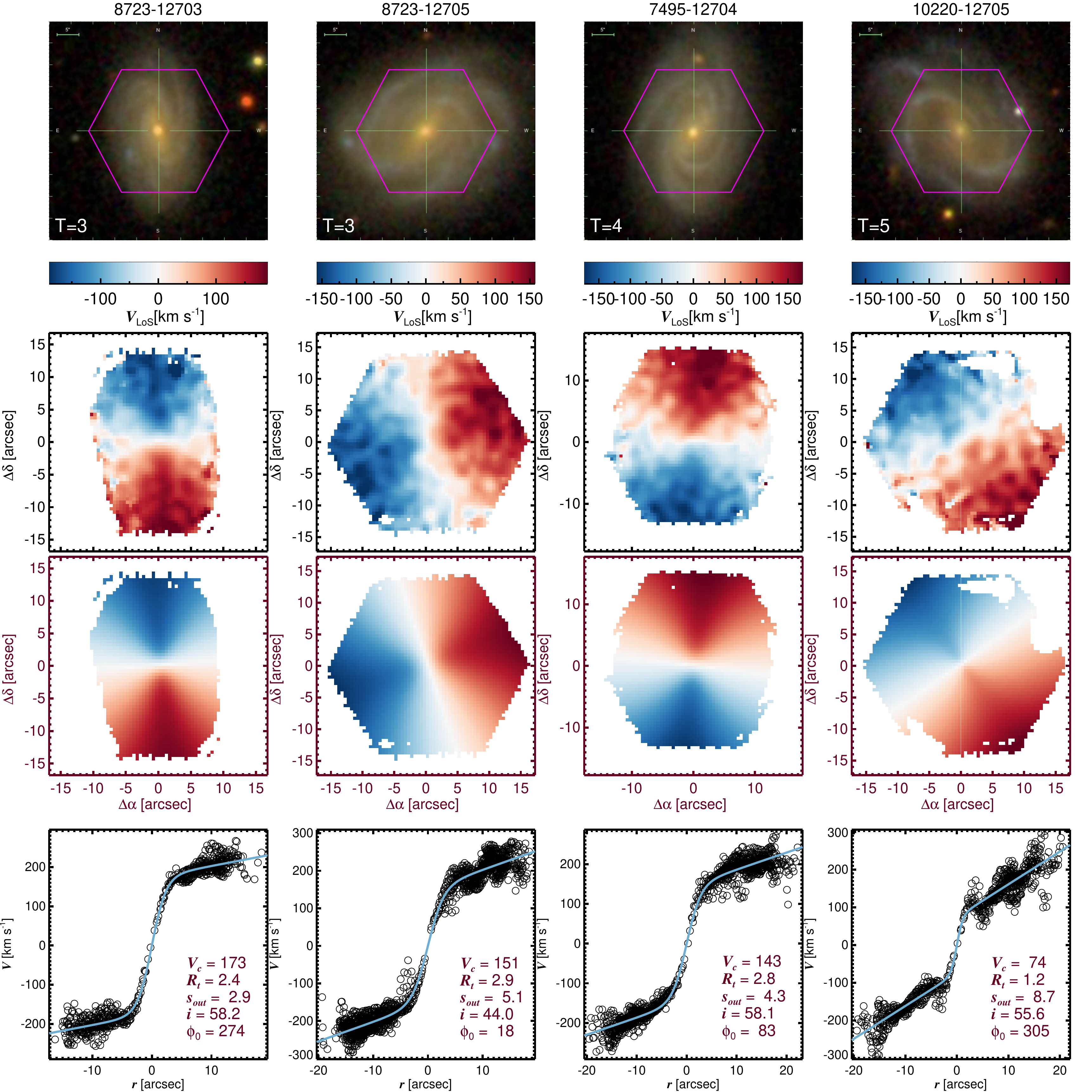}  %width=\linewidth
\centering
\caption{Examples of galaxies used in this study. These examples are galaxies with $T$-type $\ge3$. Other descriptions about the figure are identical to those in Figure \ref{fig:ex1}. The corresponding section: Section \ref{sec:sample}.
\label{fig:ex2}}
\end{figure*} 

\subsection{Extracting Line-of-sight Velocities}\label{sec:extract}
To extract line-of-sight stellar velocities from the IFU data of 10,010 galaxies, we used the code Penalized Pixel-Fitting \citep[pPXF;][]{Cappellari2004,Cappellari2017}. This code extracts the stellar kinematics and stellar population through full spectrum fitting on galaxy spectra, using the maximum penalized likelihood formalism. For model templates used in the pPXF, we used MILES single stellar population models \citep{Sanchez2006,Vazdekis2010,Falcon2011} with the Padova+00 isochrone \citep{Girardi2000} and the initial mass function of \citet{Chabrier2003}. The model templates have six metallicity values in a range of $-1.71\leq\mathrm{[M/H]}\leq0.22$ and 10 ages from 0.07 Gyr to 12.59 Gyr, so that 60 model templates were used in total. As in \citet{Belfiore2019}, we also included an eighth-order additive Legendre polynomial during the fit, in order to improve the quality of the derived stellar kinematics  (see also \citealt{Emsellem2004}). 

As in the method of \citet{Cappellari2017}, the model templates were convolved with a Gaussian function to match the resolution of the MaNGA spectra and the spectra were shifted to the rest frame before extracting stellar kinematics. The fitting was conducted after masking pixels around known emission lines and bad pixels (pixels with low coverage depth, dead fibers, or contamination from foreground stars, etc.) flagged in the MaNGA data reduction pipeline \citep{Law2016}. The fitting range was limited in a wavelength range from 3700 to 7400\AA\, in consideration of the wavelength coverage of the model templates (3540--7410\AA). The second rows in Figures \ref{fig:ex1} and \ref{fig:ex2} show extracted 2D line-of-sight velocity maps for several examples.
\\

\begin{figure}
\includegraphics[width=\linewidth]{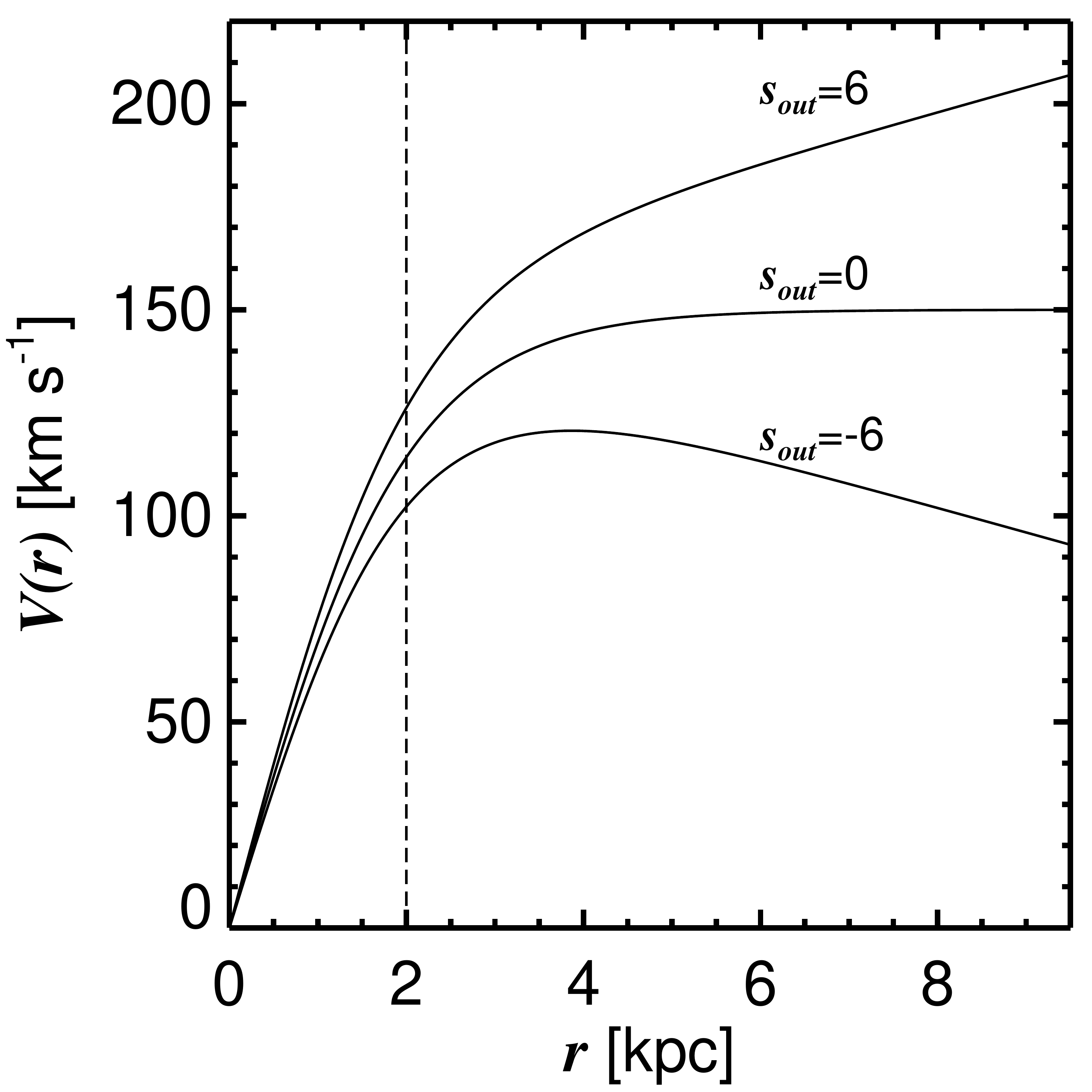}  
\centering
\caption{Examples of three rotation curves in the form of Equation \ref{eq:rc1}. All of the rotation curves have $V_c=150$ km s$^{-1}$ and $R_t=2$ kpc but have different $s_\mathrm{out}$ (6, 0, and $-6$ km s$^{-1}$kpc$^{-1}$, respectively). The vertical dashed line denotes $R_t=2$ kpc. We note that the rotation curve with $s_\mathrm{out}=-6$ km s$^{-1}$kpc$^{-1}$ does not reach $150$ km s$^{-1}$ due to the negative linear term in Equation \ref{eq:rc1}, even though $V_c$ is $150$ km s$^{-1}$. The corresponding section: Section \ref{sec:model}.
\label{fig:rcc}}
\end{figure}

\subsection{RC Model Fitting}\label{sec:model}

Here, we describe the method to fit 2D RC models to the line-of-sight velocity maps extracted in Section \ref{sec:extract}. Since our aim is to measure quantitative shapes of RCs for galaxies with diverse properties and to compare them with each other, we used an empirical RC model function that can well quantify shapes of RCs across various types of galaxies. To quantify shapes of RCs, previous studies used several functional forms such as arctangent, exponential, and hyperbolic tangent \citep{Puech2008,Feng2011,Andersen2013,Bouche2015}. RCs based on these models show a rising curve in the inner part and converge to a constant velocity in the outer part. However, many previous studies showed that RCs of galaxies have various slopes in the outer parts depending on galaxy properties \citep{Corradi1990,Casertano1991,Persic1996,Sofue2001,Noordermeer2007,Kalinova2017}. Reflecting on this fact in this study and \citet{Chung2020}, we devised a new functional form for RCs of MaNGA galaxies that is a combination of the hyperbolic tangent function and a linear term,
\begin{equation}
V(r)=V_c\tanh(r/R_t)+s_\mathrm{out}r,
\label{eq:rc1}
\end{equation}
where $s_\mathrm{out}$ is the slope of the RC at large radii $r\gg R_t$ and $R_t$ is the turnover radius where the hyperbolic tangent term begins to be flat. The coefficient of the hyperbolic tangent term $V_c$ is equal to the maximum $V(r)$ when $s_\mathrm{out}$ is 0.\footnote{The concept of the functional form for the RC used in \citet{Wiegert2014} is similar to ours in the sense that it is a combination of the hyperbolic tangent function and a linear function, although they used a different formula. See \citet{Courteau1997} for other functional forms that are more complex.} Figure \ref{fig:rcc} shows examples of three RCs from Equation \ref{eq:rc1} with different $s_\mathrm{out}$.

\begin{figure*}
\includegraphics[scale=0.55]{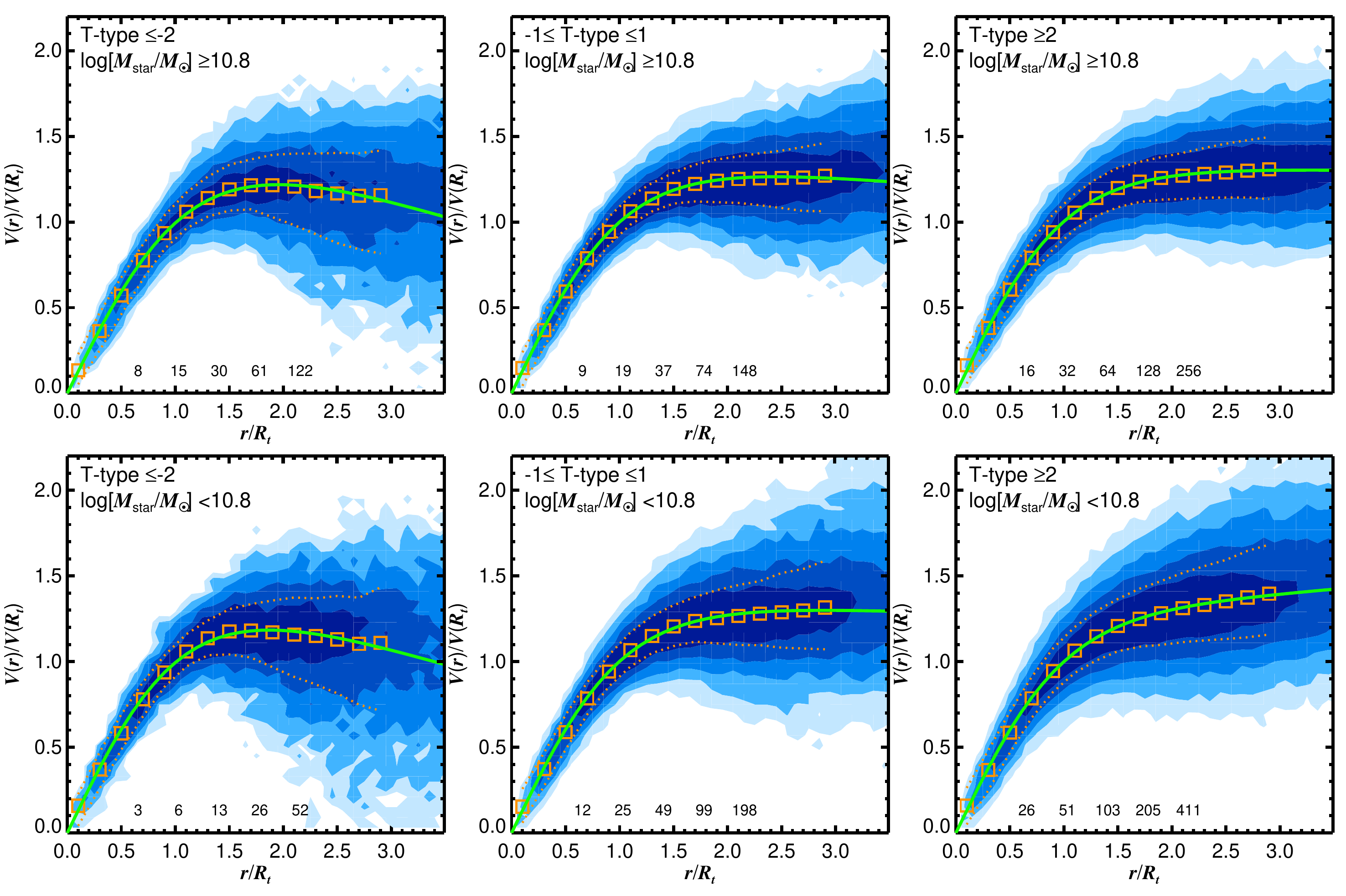}  
\centering
\caption{Stacked one-dimensional velocity profiles for galaxies in several $M_\mathrm{star}$ and $T$-type bins. In the diagrams, the distances from the kinematic center $r$ and $V(r)$ are normalized at $R_t$ and $V(R_t)$ of individual galaxies, respectively. In this figure, we only used the velocities that were derived from spectra with a median S/N$\ge10$ and located within $\pm60\degr$ from the kinematic major axes. Each velocity profile of a galaxy was folded around $r=0$ and $V(r)=0$ before stacking. The squares are median values at each bin of $r/R_t$, while the dotted lines indicate 84th and 16th percentiles of normalized velocities at the bins. The solid lines are the best-fit rotation curve models of Equation \ref{eq:rc1} fitted to the median values. The numbers at the bottom of each panel indicate the number of data points in 2D bins of 0.1 (in the $x$-axis) $\times$ 0.06 (in the $y$-axis) used for construction of the five levels of the contours. The scale of the contour levels in each panel is adjusted according to the total number of data points used in each panel. The corresponding section: Section \ref{sec:results}.
\label{fig:rcstack}} 
\end{figure*}

We note that RCs of MaNGA galaxies are well represented by the functional form of Equation \ref{eq:rc1} throughout this study. For example, individual RCs for various types of galaxies (both early and late types) properly fit into the functional form, as shown in Figures \ref{fig:ex1} and \ref{fig:ex2}. The functional form of Equation \ref{eq:rc1} also well follows median RC profiles for galaxies in diverse $M_\mathrm{star}$ and morphology bins (Figure \ref{fig:rcstack}). The fact that the same functional form can be applied to both early- and late-type galaxies is consistent with recent studies of stellar kinematics of galaxies that revealed that a large fraction of elliptical galaxies, even those with apparently round and nondisky shapes, contain substantial rotating disk components like late-type galaxies \citep{Cappellari2016,Graham2018}. Furthermore, RCs of simple models consisting of a baryonic disk and a dark matter halo can also be described by this functional form (Figures \ref{fig:m_re}, \ref{fig:m_bry}, \ref{fig:m_dm}, and \ref{fig:m_vd}). 

The radial coverage of RCs for MaNGA galaxies in the final sample used here is $\sim5R_t$ on average. Thus, the functional form of Equation \ref{eq:rc1} can be regarded as a good RC model within this range, although it may not be suitable at very large radii \citep{Noordermeer2007}, for which extraction of stellar kinematics using MaNGA data is nearly impossible.

To fit our RC model on the 2D line-of-sight velocity maps extracted in Section \ref{sec:extract}, we used a 2D model \citep{Beckman2004,Chung2020} such as
\begin{equation}
V_\mathrm{obs}(r', \phi')=V_\mathrm{sys}+V(r)\sin i\cos(\phi-\phi_0).
\label{eq:rc2}
\end{equation}
In this equation, $V_\mathrm{obs}(r', \phi')$ is the observed 2D line-of-sight velocity map where $r'$ is the radial distance from the kinematic center of a galaxy to each pixel on the plane of the sky, and $\phi'$ is the angle of the pixel. Parameters $r$ and $\phi$ are the radial coordinate and angle in the plane of the galaxy (deprojected plane), respectively. $V(r)$ in this equation is the adopted RC function of Equation \ref{eq:rc1}. Parameter $i$ is the kinematic inclination angle and $\phi_0$ is the kinematic position angle of the galaxy. $V_\mathrm{sys}$ is the systematic line-of-sight velocity at the kinematic center. Including the kinematic center position ($x_\mathrm{center}$ and $y_\mathrm{center}$) in the plane of the sky, eight fitting parameters are used in the fitting process ($x_\mathrm{center}$, $y_\mathrm{center}$, $V_\mathrm{sys}$, $i$, $\phi_0$, $V_c$, $R_t$, and $s_\mathrm{out}$).

We used the minimum $\chi^2$ method for the 2D fitting, finding a set of parameters that minimizes $\chi^2$ between the observed line-of-sight map and the 2D model of Equation \ref{eq:rc2}.  As in Section \ref{sec:extract}, pixels contaminated by foreground stars were masked in the fitting. In the fitting, we only used line-of-sight velocities derived from spectra with a median S/N larger than or equal to 5. 

The third rows in Figures \ref{fig:ex1} and \ref{fig:ex2} show several examples of best-fit 2D models (Equation \ref{eq:rc2}) for observed velocity maps that are in the second rows in the figures. The bottom rows in the same figures show observed one-dimensional velocity profiles with the best-fit RC models of Equation \ref{eq:rc1}.

Through intensive tests with mock IFU data, \citet{Chung2020} found that the maximum radial coverage of the 2D velocity maps along the major axis ($R_\mathrm{max}$) should be at least larger than $\sim2.5R_t$ (or conservatively $\sim3.0R_t$) in order to robustly determine the outer slope of the RC ($s_\mathrm{out}$). For this reason, we only used galaxies with $R_\mathrm{max}/R_t \geq 3$.\footnote{In this study, $R_\mathrm{max}$ was calculated using pixels in the velocity map that are within $\pm5\degr$ from the major axis and correspond to spectra with a median S/N $\ge5$.} To exclude IFU data that cover only a small (or central) part of galaxies, we ruled out galaxies whose $R_\mathrm{max}$ is smaller than $R_e$. Applying these radial coverage criteria, the number of MaNGA galaxies is 2829. 

RCs for galaxies with low kinematic inclinations (close to face-on) can be uncertain due to the strong coupling between the $\sin i$ and $V_c$ terms in Equations \ref{eq:rc1} and \ref{eq:rc2} \citep{Noordermeer2007,Chung2020}. A test in \citet{Chung2020}  reveals that RCs can be reliably determined for kinematic inclinations of $i\gtrsim20\degr$ -- $25\degr$. Therefore, we excluded galaxies with $i<20\degr$ in this study. The number of MaNGA galaxies after this cut is 2498. As mentioned in Section \ref{sec:morph}, we excluded edge-on galaxies, so that $99\%$ of galaxies in the final sample have $i<75\degr$.

For a test, we converted the kinematic inclination angles of galaxies in the final sample into axis ratios of rotating disks ($q_k$), assuming that the rotation components in galaxies are thin disks with intrinsic axis ratios of 1, and then compared them with photometric axis ratios ($q_p$) estimated from a 2D S{\'e}rsic model in the NSA catalog. By doing so, we found that $q_k$ is very consistent with $q_p$, so that the average difference between them is only 0.006 and the standard deviation of the difference is 0.15, which implies that the kinematic inclination angles in this study were not measured erroneously.
\\

\begin{deluxetable*}{c ccccc ccccc cccc}
\tablecaption{$T$-type Classification Scheme \label{tbm}}
\tabletypesize{\scriptsize}
\tablehead{\colhead{Class} & \colhead{E} & \colhead{E/S0} & \colhead{S0} & \colhead{S0/a} & \colhead{Sa} & \colhead{Sab} & \colhead{Sb} & \colhead{Sbc} & \colhead{Sc} & \colhead{Scd} & \colhead{Sd} & \colhead{Sdm} & \colhead{Sm} & \colhead{Im} 
}
\startdata
$T$-type & -3 & -2 & -1 & 0 & 1 & 2 & 3 & 4 & 5 & 6 & 7 & 8 & 9 & 10\\
\enddata
\end{deluxetable*}

\begin{figure}
\includegraphics[width=\linewidth]{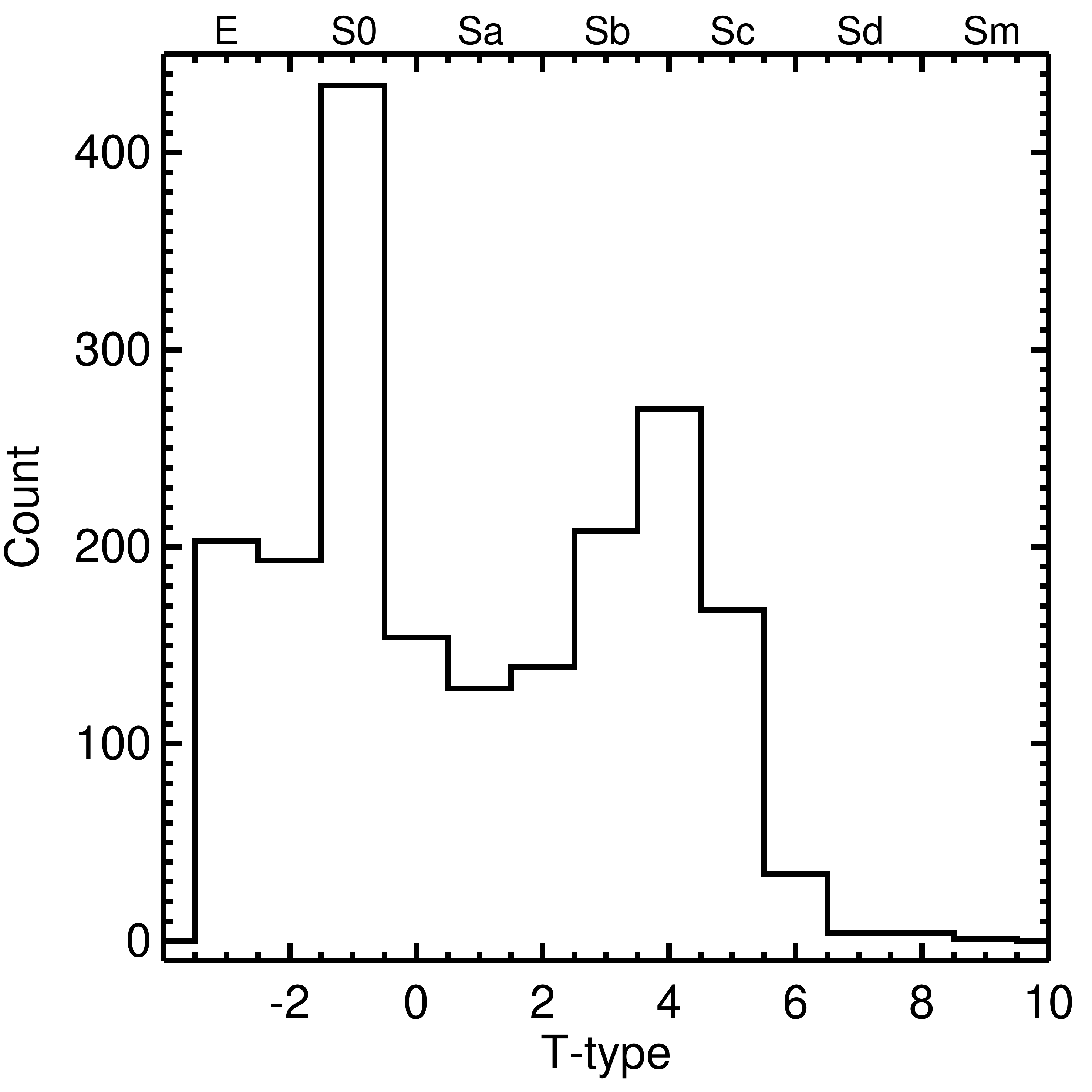}
\centering
\caption{$T$-type distribution for the final sample. The corresponding section: Section \ref{sec:morph}.
\label{fig:ttyped}}
\end{figure}

\begin{figure*}
\includegraphics[scale=0.295]{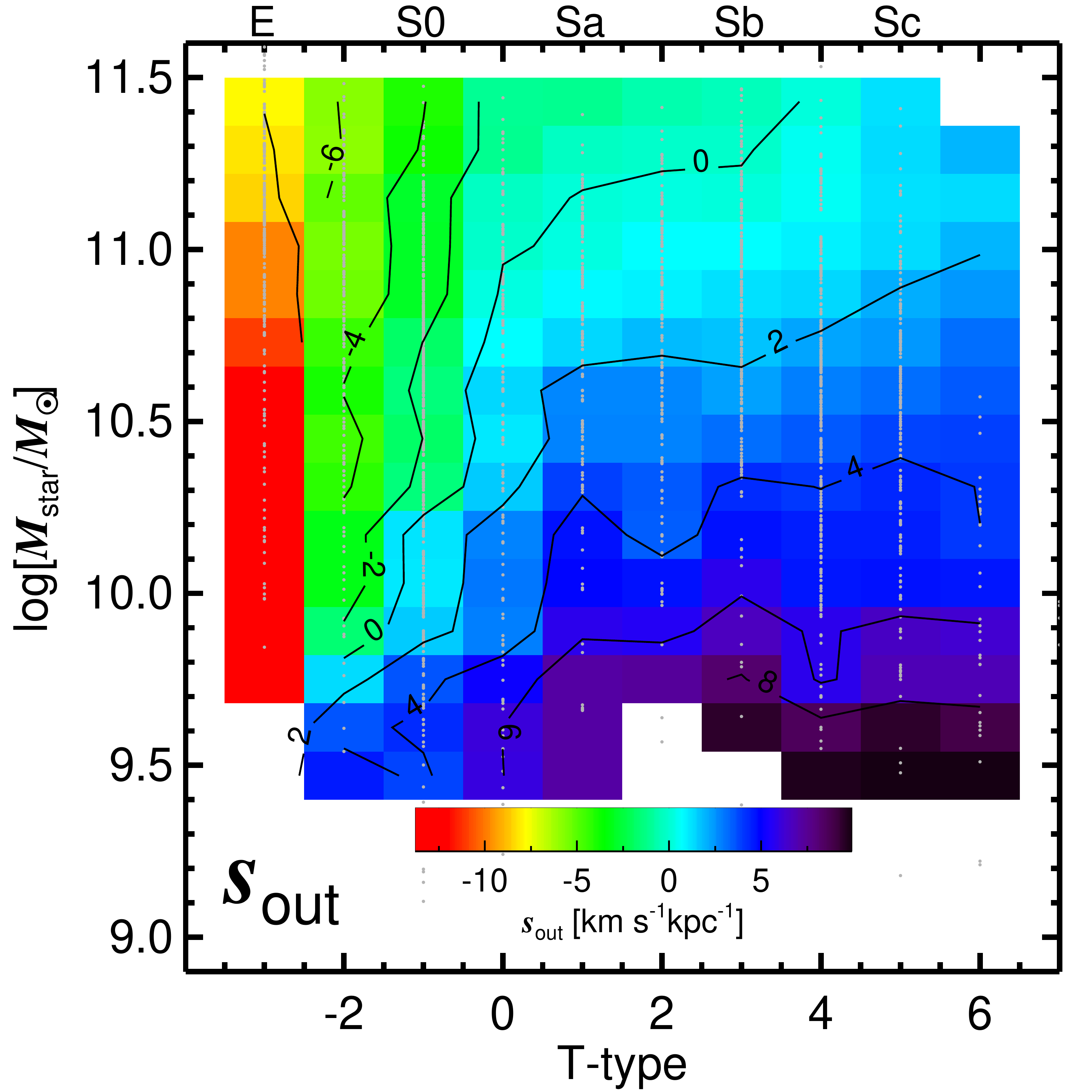} \includegraphics[scale=0.295]{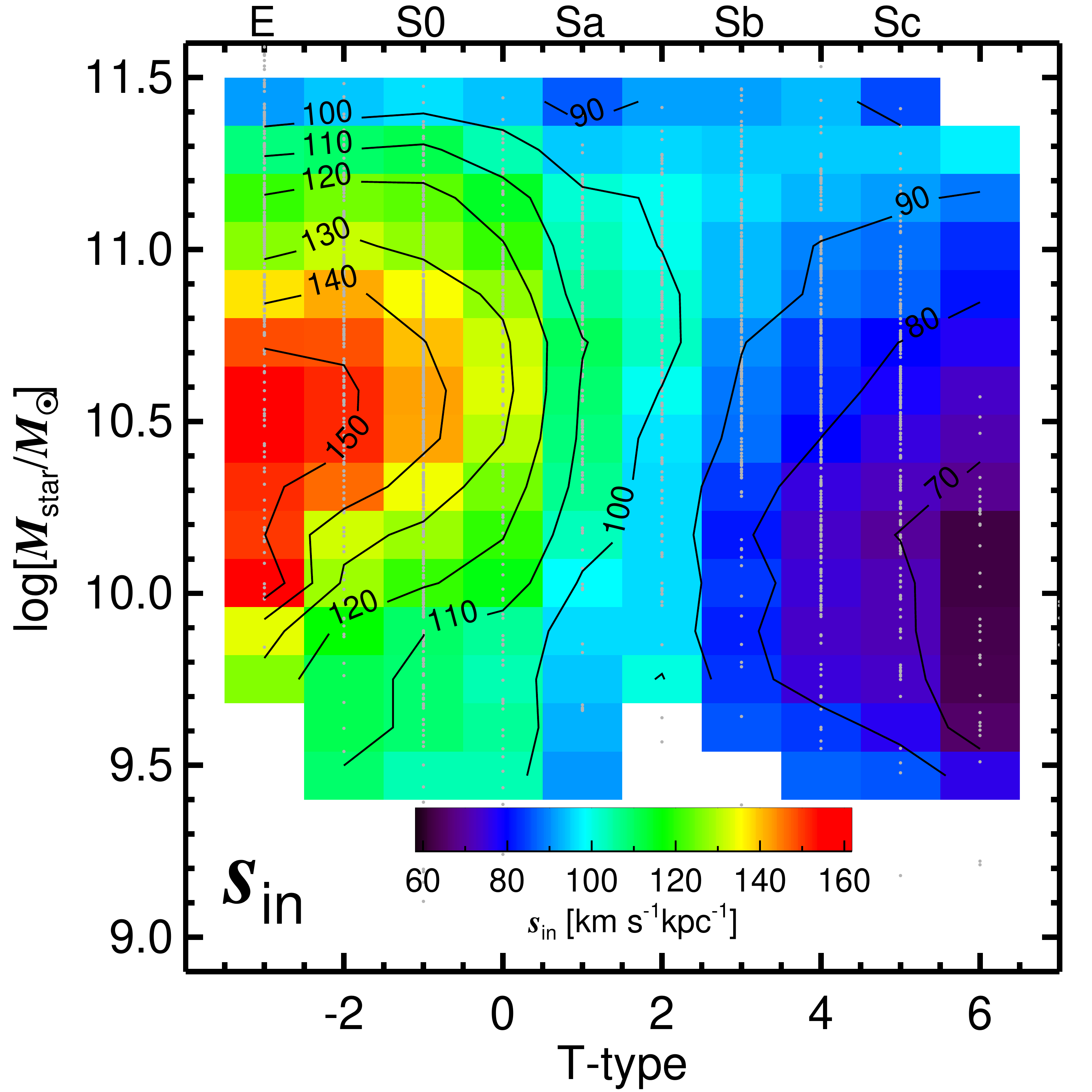}  
\centering
\caption{Inner and outer slopes of rotation curves ($s_\mathrm{in}$ and $s_\mathrm{out}$) in the $T$-type versus $\log M_\mathrm{star}$ plane. The left panel is for $s_\mathrm{out}$ and the right panel is for $s_\mathrm{in}$. The gray dots are individual galaxies. Slopes $s_\mathrm{out}$ and $s_\mathrm{in}$ are represented by colors (see the color bars for the color-coded $s_\mathrm{out}$ and $s_\mathrm{in}$) and contours (values in the middle of contour lines indicate $s_\mathrm{out}$ or $s_\mathrm{in}$). To construct color maps and contours for $s_\mathrm{out}$ and $s_\mathrm{in}$, we used a grid in which block sizes in the $T$-type and $\log M_\mathrm{star}$ axes are 1 and 0.14, respectively. At each point in the grid, we calculated the median slopes for galaxies within a rectangular bin whose sizes in the $T$-type and $\log M_\mathrm{star}$ sides are 2 and 0.7, respectively. We only used the 2D bins in which the number of galaxies is larger than 20. The corresponding section: Section \ref{sec:results}.
\label{fig:slope}}
\end{figure*}

\begin{figure*}
\includegraphics[width=\linewidth]{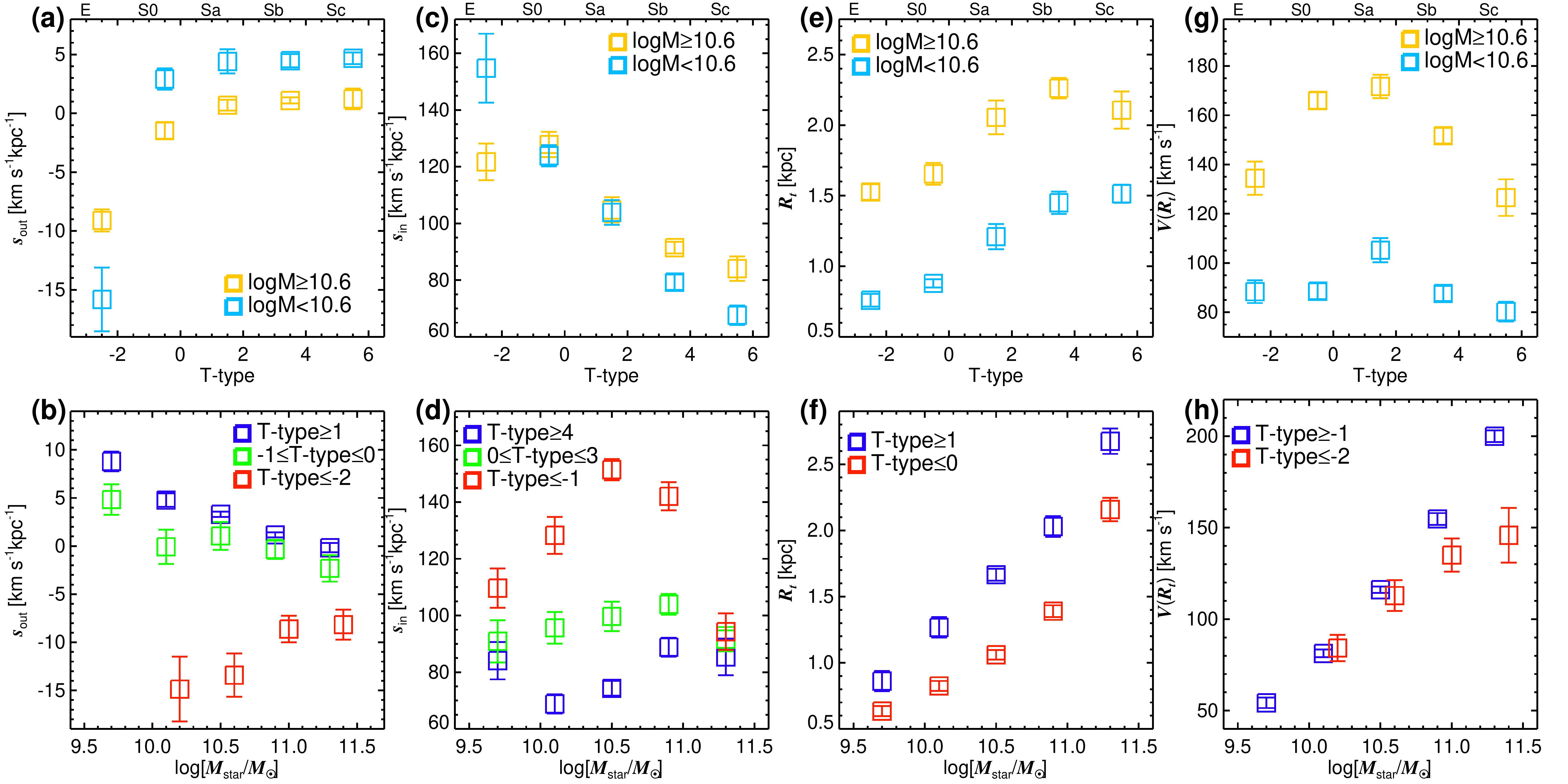}
\centering
\caption{Rotation curve parameters as a function of $T$-type (the upper panels) and  $\log M_\mathrm{star}$ (the lower panels). Panels (a) and (b): $s_\mathrm{out}$; (c) and (d): $s_\mathrm{in}$; (e) and (f): $R_t$; (g) and (h): $V(R_t)$. The square denotes the median value within a bin whose size is 2 and 0.4 in the upper and lower panels, respectively. The error bar is the standard deviation of the median values from 1000 bootstrap resamplings. Galaxies are divided into two identical mass bins in the upper panels. By contrast, in the lower panels, we divide galaxies into two or three different $T$-type bins, in order to clearly display different trends in different $T$-type classes. The corresponding section: Section \ref{sec:results}.
\label{fig:aplot}}
\end{figure*}

\subsection{Morphology Classification}\label{sec:morph}

We assigned $T$-type numbers to galaxies according to their morphological classes. The $T$-type system is a widely used morphology classification scheme \citep{Nair2010,Ann2015,Dominguez2018}. We used 14 morphological classes and hence 14 $T$-type values from $-3$ to $10$ as shown in Table \ref{tbm}. Galaxies with $T$-type $<0$ are early types (elliptical and lenticular galaxies), while those with $T$-type $>0$ correspond to late types (from spiral to irregular galaxies). Major morphological classes (e.g., E, S0, Sa, Sb, etc.) are odd numbers in our $T$-type scheme, whereas intermediate classes between major ones (e.g., S0/a, Sab, Sbc, etc.) are even numbers.

We classified $T$-types of galaxies by visually inspecting combined color images of the $g$, $r$, and $i$ bands in SDSS (see the first rows of Figures \ref{fig:ex1} and \ref{fig:ex2}). A morphological characteristic of elliptical galaxies (E, $T$-type $=-3$) is a smooth, featureless light distribution with an extended stellar halo in the outside of galaxies. Lenticular galaxies (S0, $T$-type $=-1$) have a large and bright bulge surrounded by a disk-like structure with no visible spiral arms (sometimes they have a ring structure). Sa galaxies ($T$-type $=1$) show a bright bulge and very tightly wound smooth arms. Sb galaxies ($T$-type $=3$) have a fainter bulge and less tight spiral arms than Sa. Their spiral arms are usually blue and prominent. Sc galaxies ($T$-type $=5$) are marked by a small and/or faint bulge with loosely wound blue arms. Resolved stellar clusters and nebulae can be seen in some Sc galaxies. Sd galaxies ($T$-type $=7$) have a very faint bulge with a blue disk in which spiral arms are generally fragmented into stellar clusters and nebulae. Galaxies with irregular morphologies are classified into Sm ($T$-type $=9$) or Im ($T$-type $=10$). Among them, Sm galaxies show weak spiral structures. 

In the visual inspection, we found that 515 galaxies are too inclined to determine their morphology correctly. We also visually inspected results of the RC model fitting (Section \ref{sec:model}) and found that 43 galaxies have bad fitting qualities. We excluded those edge-on (too inclined) galaxies and galaxies with bad fitting qualities in the final sample. Thus, the number of galaxies in our final sample is 1940. We note that the number of early-type galaxies with $T$-type $<0$ is 830, while that of late-type galaxies with $T$-type $>0$ is 956 (the number of galaxies with T $=0$ is 154). Most of galaxies in the final sample ($99.5\%$) have $T$-type $\le6$. See Figure \ref{fig:ttyped} for the $T$-type distribution for our final sample.

In order to check the reliability of our $T$-type classifications, we compared our classifications with those in two other catalogs. The first catalog is a value-added catalog of SDSS containing visually classified morphology information for all MaNGA galaxies in SDSS Data Release (DR) 15 (4614 galaxies). The visual inspection of morphology was conducted based on images of SDSS and the DESI (Dark Energy Spectroscopic Instrument) Legacy Imaging Survey \citep{Dey2019} following the methods in \citet{Hernandez2010}. A total of 846 galaxies in our sample are also in that catalog. Since the $T$-type scheme for early-type galaxies ($T$-type $<0$) in the catalog is slightly different from ours,\footnote{$T$-type $=-5$ for E and $T$-type $=-2$ for S0, with essentially no intermediate type between them.}  we only conducted a comparison using our late-type galaxies with $T$-type $>0$ and found that the mean difference of $T$-type is $-0.19\pm0.71$\footnote{The error is the standard deviation of the difference.} after excluding $3\sigma$ outliers (our $T$-type values are slightly smaller on average). The mean difference is reduced to $-0.01\pm0.55$ for galaxies with $T$-type $\ge3$ in our sample .

The second catalog for comparison is a catalog of \citet{Dominguez2018} that has morphology information for SDSS DR7 galaxies. The classifications of the catalog were obtained with deep learning algorithms using convolutional neural networks. For the training of the $T$-type classification, they used a morphology catalog of \citet{Nair2010}, in which $T$-type classifications were determined for SDSS DR4 galaxies based on visual inspection. The catalog of \citet{Dominguez2018} uses the same $T$-type scheme as ours, except that the $T$-type values in their catalog are non-integer numbers (e.g., $T$-type $=2.3$) unlike our $T$-type values that are all integers. A total of 1646 galaxies in our sample are also in this catalog. We found that the mean difference in $T$-type between our study and the catalog of \citet{Dominguez2018} is $0.07\pm1.11$ after excluding $3\sigma$ outliers. For galaxies with $T$-type $<0$ in our sample, the mean difference is $0.13\pm0.91$,\footnote{Galaxies with T$=-3$ and T$=-1$ in our classification have on average T$=-2.4$ and T$=-1.5$, respectively, in the catalog of \citet{Dominguez2018}.} while it is $0.04\pm1.15$ for galaxies with $T$-type $>0$. Through the comparison of the two catalogs, we conclude that our $T$-type classifications are consistent with those in the other two catalogs within about $\pm1$. 

We selected 1940 galaxies for the final sample with various criteria described above. The selection work excludes galaxies evenly in the IFU bundle size and (near-ultraviolet  -- $i$) color distributions of the whole MaNGA galaxy sample without any bias. The selection of the final sample cuts low-mass galaxies a little bit more, so that the median $M_\mathrm{star}$ of the final sample is $10^{10.7}M_\odot$, while that of the whole MaNGA galaxy sample is $10^{10.5}M_\odot$. In conclusion, the final sample used in this study still represents the whole MaNGA galaxy sample in terms of (near-ultraviolet  -- $i$) color, $M_\mathrm{star}$, and the IFU bundle size.
\\

\begin{figure}
\includegraphics[width=\linewidth]{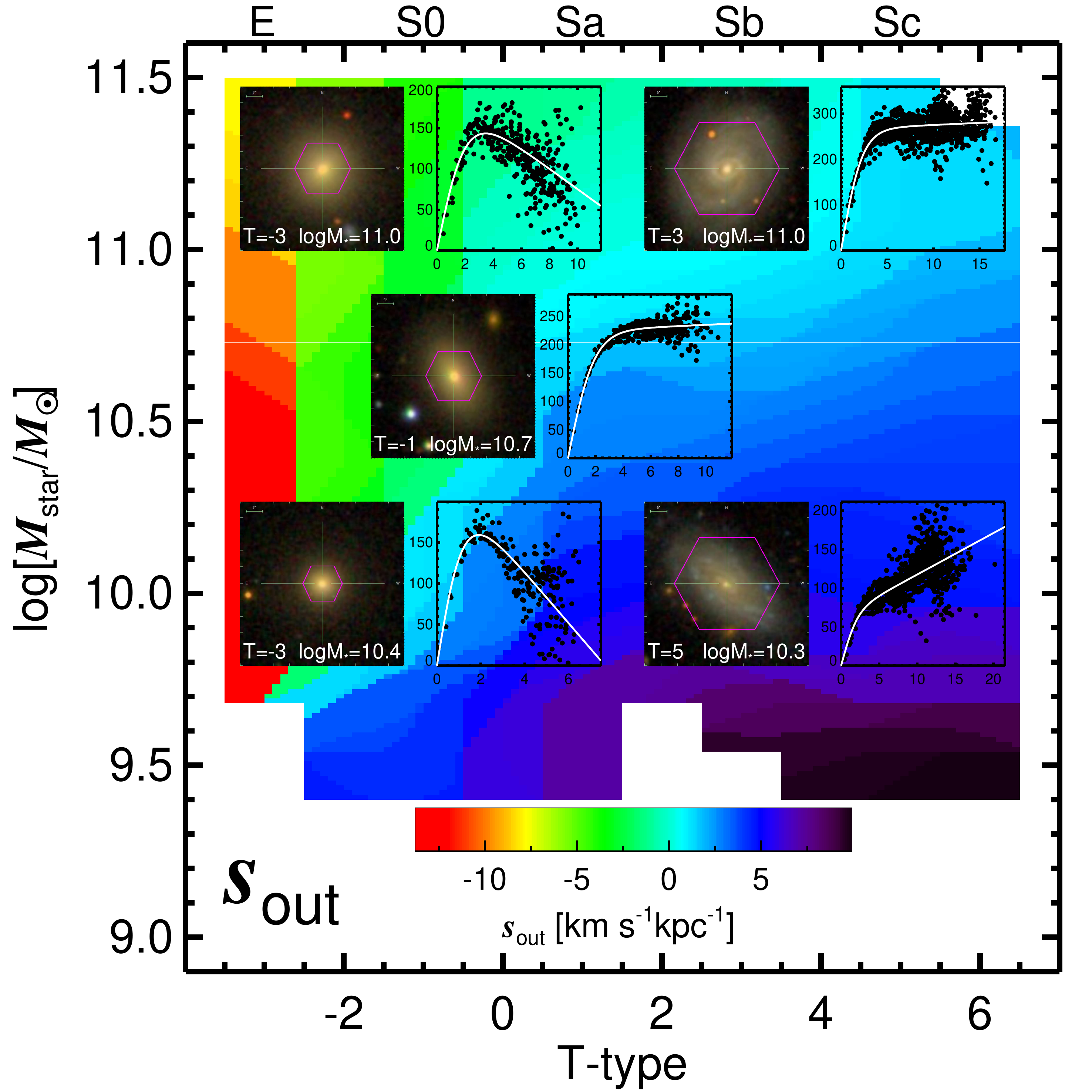}
\centering
\caption{A color map showing $s_\mathrm{out}$ in the $T$-type versus $\log M_\mathrm{star}$ plane, which is a smoothed version of that in the left panel of Figure \ref{fig:slope}. Thumbnail images and one-dimensional velocity profiles for five typical examples are superposed. The description for the one-dimensional velocity profiles and the best-fit rotation curve models (the white solid lines) are the same as that in Figure \ref{fig:ex1}, except that they are folded around $r=0$ and $V(r)=0$ here. The superposed figures are placed in the map according to the $T$-type and $M_\mathrm{star}$ values of the galaxies, which are marked in the bottoms of the thumbnail images. The corresponding section: Section \ref{sec:results}.
\label{fig:slope_ex}}
\end{figure}

\section{Results}\label{sec:results}
Here, we present how the RC parameters in Equation \ref{eq:rc1} depend on $M_\mathrm{star}$ and $T$-type. Firstly, we examined the inner and outer slopes of RCs. The outer slope is $s_\mathrm{out}$ in Equation \ref{eq:rc1}. The inner slope is defined as 
\begin{equation}
s_\mathrm{in}=\lim_{r\to0}\frac{V(r)}{r},
\label{eq:sin}
\end{equation}
where $V(r)$ is the RC model function of Equation \ref{eq:rc1}. Thus, $s_\mathrm{in}$ is a slope of the tangent line at the center of the RC \citep{Lelli2013,Erroz2016}.

The left panel of Figure \ref{fig:slope} shows $s_\mathrm{out}$ in the $T$-type versus $\log M_\mathrm{star}$ plane, and panels (a) and (b) of Figure \ref{fig:aplot} display median $s_\mathrm{out}$ as a function of $T$-type and $\log M_\mathrm{star}$, respectively. For late-type galaxies with $T$-type $\ge1$, galaxies with lower $M_\mathrm{star}$ have positively steeper $s_\mathrm{out}$, so that $s_\mathrm{out}$ changes from $0$ to $9$ km s$^{-1}$kpc$^{-1}$ ($\Delta s_\mathrm{out}=9$ km s$^{-1}$kpc$^{-1}$) when $M_\mathrm{star}$ goes from $10^{11.5}$ to $10^{9.5} M_\odot$. By contrast, $s_\mathrm{out}$ only changes less than $1$ km s$^{-1}$kpc$^{-1}$ at a given $M_\mathrm{star}$ in a range of $1\le$ $T$-type $\le6$. 

Early-type galaxies with $T$-type $\le0$ have smaller $s_\mathrm{out}$ at smaller $T$-type. This trend is more severe for the lower-mass range of  $\log(M_\mathrm{star}/M_\odot)<10.7$ ($\Delta s_\mathrm{out}>17$ km s$^{-1}$kpc$^{-1}$) than for the higher-mass range of $\log(M_\mathrm{star}/M_\odot)\ge10.7$ ($\Delta s_\mathrm{out}=10$ km s$^{-1}$kpc$^{-1}$). Consequently, elliptical galaxies with $T$-type $\le-2$ have negative $s_\mathrm{out}$ and their $s_\mathrm{out}$ falls down as $M_\mathrm{star}$ decreases ($\Delta s_\mathrm{out}=7$ km s$^{-1}$kpc$^{-1}$ in $10^{10.0}<M_\mathrm{star}/M_\odot<10^{11.6}$), which is the opposite trend to $s_\mathrm{out}$ for $T$-type $\ge1$. 

The results for $s_\mathrm{out}$ can also be checked in Figure \ref{fig:rcstack} showing stacked one-dimensional velocity profiles (normalized at $R_t$ and $V(R_t)$) for galaxies in several $M_\mathrm{star}$ and $T$-type bins. In Figure \ref{fig:slope_ex}, we also show thumbnail images and one-dimensional velocity profiles for typical galaxies that are superposed on the color map of $s_\mathrm{out}$ in the $T$-type versus $\log M_\mathrm{star}$ plane.

The right panel of Figure \ref{fig:slope} shows $s_\mathrm{in}$ in the $T$-type versus $\log M_\mathrm{star}$ plane, and panels (c) and (d) of Figure \ref{fig:aplot} display median $s_\mathrm{in}$ as a function of $T$-type and $\log M_\mathrm{star}$, respectively. In general, $s_\mathrm{in}$ is a strong function of $T$-type in such a way that galaxies with a smaller $T$-type have a steeper $s_\mathrm{in}$ ($\sim130$ km s$^{-1}$kpc$^{-1}$ at $T$-type $\le-2$) than those with a larger $T$-type ($s_\mathrm{in}\sim70$ km s$^{-1}$kpc$^{-1}$ at $T$-type $\ge5$). The $M_\mathrm{star}$ dependence is also seen in $s_\mathrm{in}$. For $T$-type $\ge3$,  $s_\mathrm{in}$ tends to mildly increase with $M_\mathrm{star}$ (from $70$ to $90$ km s$^{-1}$kpc$^{-1}$ in $10^{10.0}<M_\mathrm{star}/M_\odot<10^{11.5}$). By contrast, for $T$-type $\le-1$, the maximum $s_\mathrm{in}$ appears at $\log(M_\mathrm{star}/M_\odot)\approx10.5$ ($\sim150$ km s$^{-1}$kpc$^{-1}$) and less massive and more massive galaxies than $\log(M_\mathrm{star}/M_\odot)\approx10.5$ have a smaller $s_\mathrm{in}$. The most massive early-type galaxies with $T$-type $\le-1$ and $\log(M_\mathrm{star}/M_\odot)\gtrsim11.3$ have $s_\mathrm{in}<100$ km s$^{-1}$kpc$^{-1}$. 

\begin{figure*}
\includegraphics[scale=0.295]{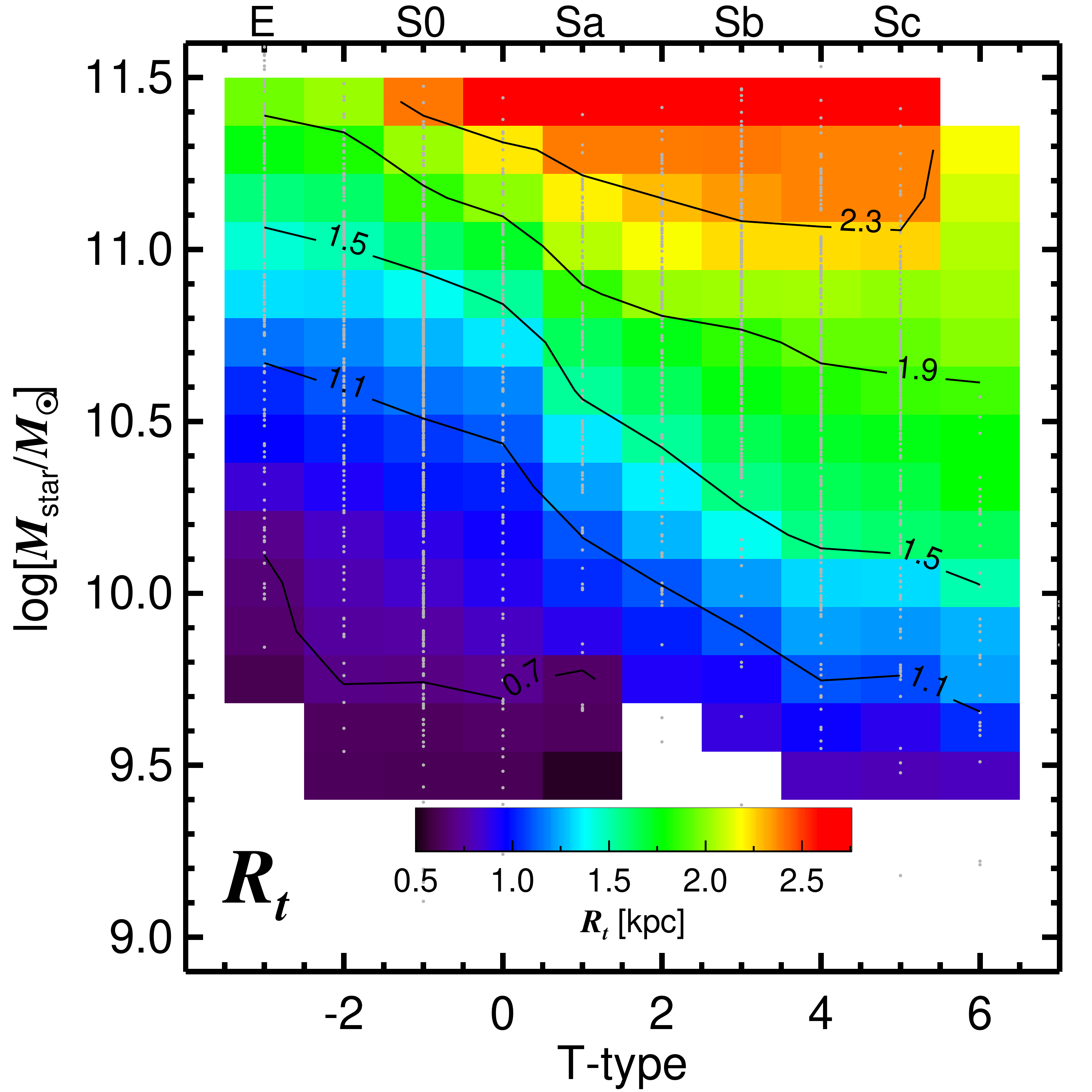} \includegraphics[scale=0.295]{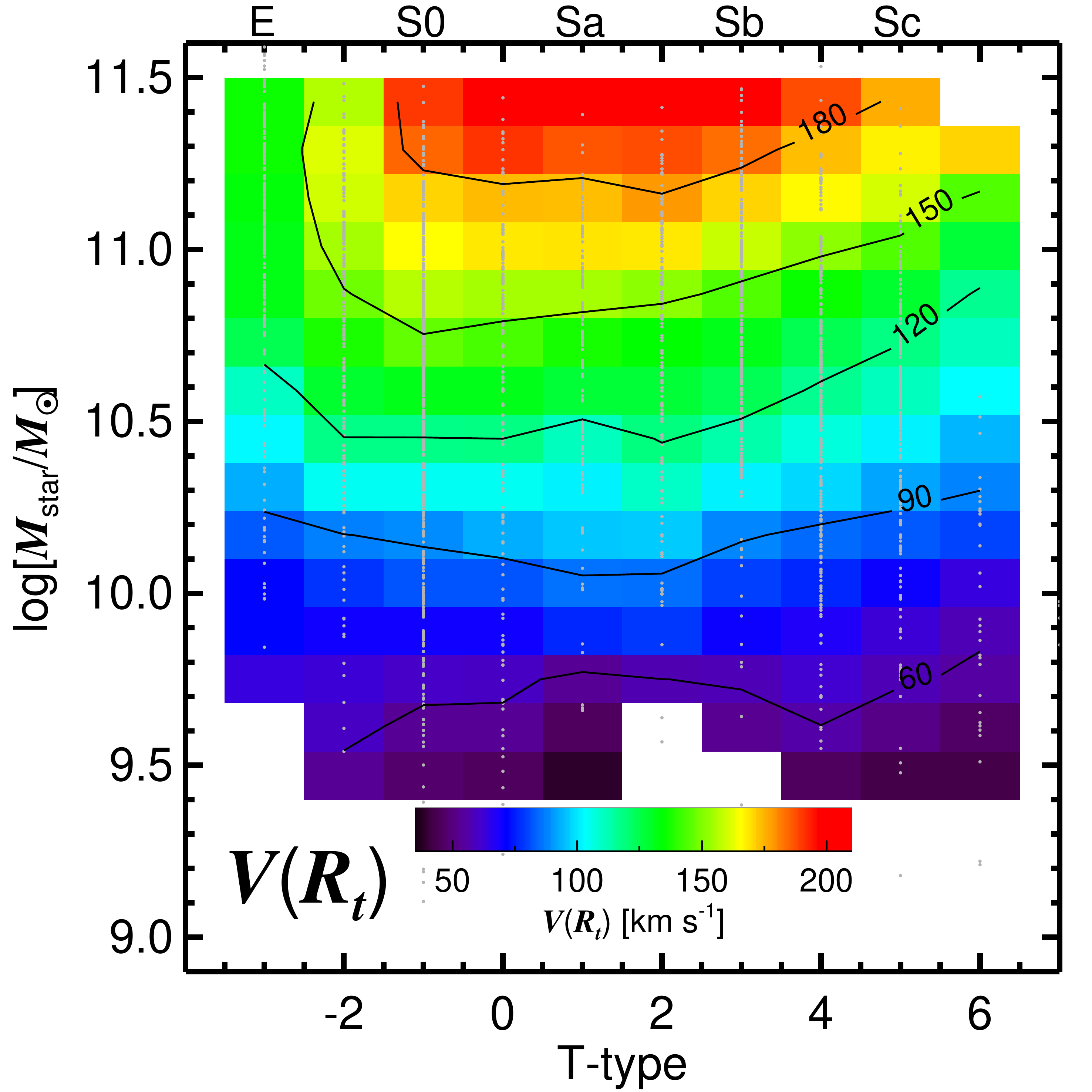}  
\centering
\caption{$R_t$ and $V(R_t)$ in the $T$-type versus $\log M_\mathrm{star}$ plane. The left panel is for $R_t$ and the right panel is for $V(R_t)$. $R_t$ and $V(R_t)$ are represented by colors (see the color bars for the color-coded $R_t$ and $V(R_t)$) and contours (values in the middle of contour lines indicate $R_t$ and $V(R_t)$). Other descriptions about the construction of the color maps and contours are identical to those in Figure \ref{fig:slope}. The corresponding section: Section \ref{sec:results}.
\label{fig:rv}}
\end{figure*} 

\begin{figure}
\includegraphics[width=\linewidth]{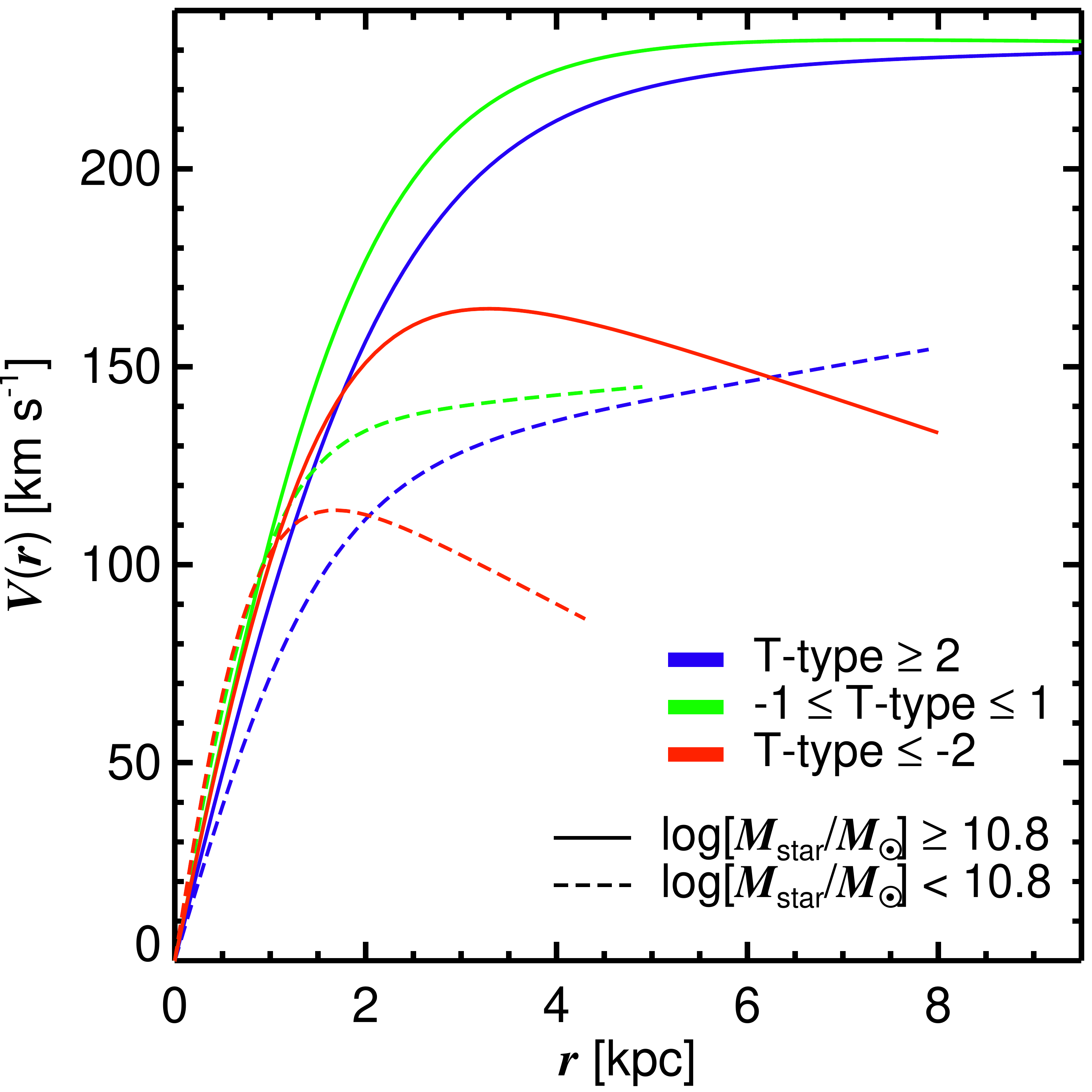}  
\centering
\caption{Typical rotation curves of galaxies in various $T$-type and $M_\mathrm{star}$ bins. These rotation curves were constructed by taking median values of rotation curve parameters (Equation \ref{eq:rc1}) in each bin. The rotation curve profiles are displayed up to $5R_t$. The corresponding section: Section \ref{sec:results}.
\label{fig:rcbin}}
\end{figure} 

Next, we investigated $R_t$ and the rotation velocity at $R_t$ ($V(R_t)$), which can be regarded as the typical rotation velocity of a galaxy. The left panel of Figure \ref{fig:rv} shows $R_t$ in the $T$-type versus $\log M_\mathrm{star}$ plane, and panels (e) and (f) of Figure \ref{fig:aplot} display median $R_t$ as a function of $T$-type and $\log M_\mathrm{star}$, respectively. $R_t$ is well correlated with $M_\mathrm{star}$ ($\Delta R_t\sim1.7$ kpc from $10^{9.5}$ to $10^{11.5} M_\odot$) in the sense that galaxies with a higher $M_\mathrm{star}$ have a larger $R_t$. The linear Pearson correlation coefficient ($C_P$) of $R_t$ and $\log M_\mathrm{star}$ is $0.53$. $R_t$ also depends on $T$-type in the sense that galaxies with a smaller $T$-type have a smaller $R_t$ at a given $M_\mathrm{star}$ ($\Delta R_t\sim0.8$ kpc from $T$-type $=-3$ to $5$). The reason for the $T$-type dependence is that $R_t$ also depends on $R_e$ (see Section \ref{sec:discuss}) in such a way that galaxies with a smaller $R_e$ have a smaller $R_t$ ($C_P=0.64$ for $R_t$ and $R_e$) and earlier-type galaxies (smaller $T$-type) have a smaller $R_e$ at a given $M_\mathrm{star}$ \citep{Shen2003}.

The right panel of Figure \ref{fig:rv} shows $V(R_t)$ in the $T$-type versus $\log M_\mathrm{star}$ plane, and panels (g) and (h) of Figure \ref{fig:aplot} display median $V(R_t)$ as a function of $T$-type and $\log M_\mathrm{star}$, respectively. $V(R_t)$ mainly correlates with $M_\mathrm{star}$ so that galaxies with a higher $M_\mathrm{star}$ have a larger $V(R_t)$ ($C_P=0.59$ for $\log M_\mathrm{star}$ and $V(R_t)$). We note that massive elliptical galaxies with $T$-type $\le-2$ and $\log(M_\mathrm{star}/M_\odot)\ge11.0$ have a noticeably smaller $V(R_t)$ ($\sim140$ km s$^{-1}$) than galaxies with $T$-type $>-1$ with similar $M_\mathrm{star}$ ($V(R_t)\sim190$ km s$^{-1}$). This is because such massive elliptical galaxies are generally not rotation-supported systems, in contrast to disk galaxies \citep{Emsellem2011,Penoyre2017,Veale2017,Graham2018,Lagos2018}. 

The results for $s_\mathrm{out}$, $s_\mathrm{in}$, $R_t$, and $V(R_t)$ can also be identified in Figure \ref{fig:rcbin}. Figure \ref{fig:rcbin} shows typical RCs of galaxies in various $T$-type and $M_\mathrm{star}$ bins, which were constructed by taking median values of RC parameters in each bin.  
\\

\section{Discussion: What Determines the Shapes of RC\lowercase{s}?}\label{sec:discuss}

\begin{figure}
\includegraphics[width=\linewidth]{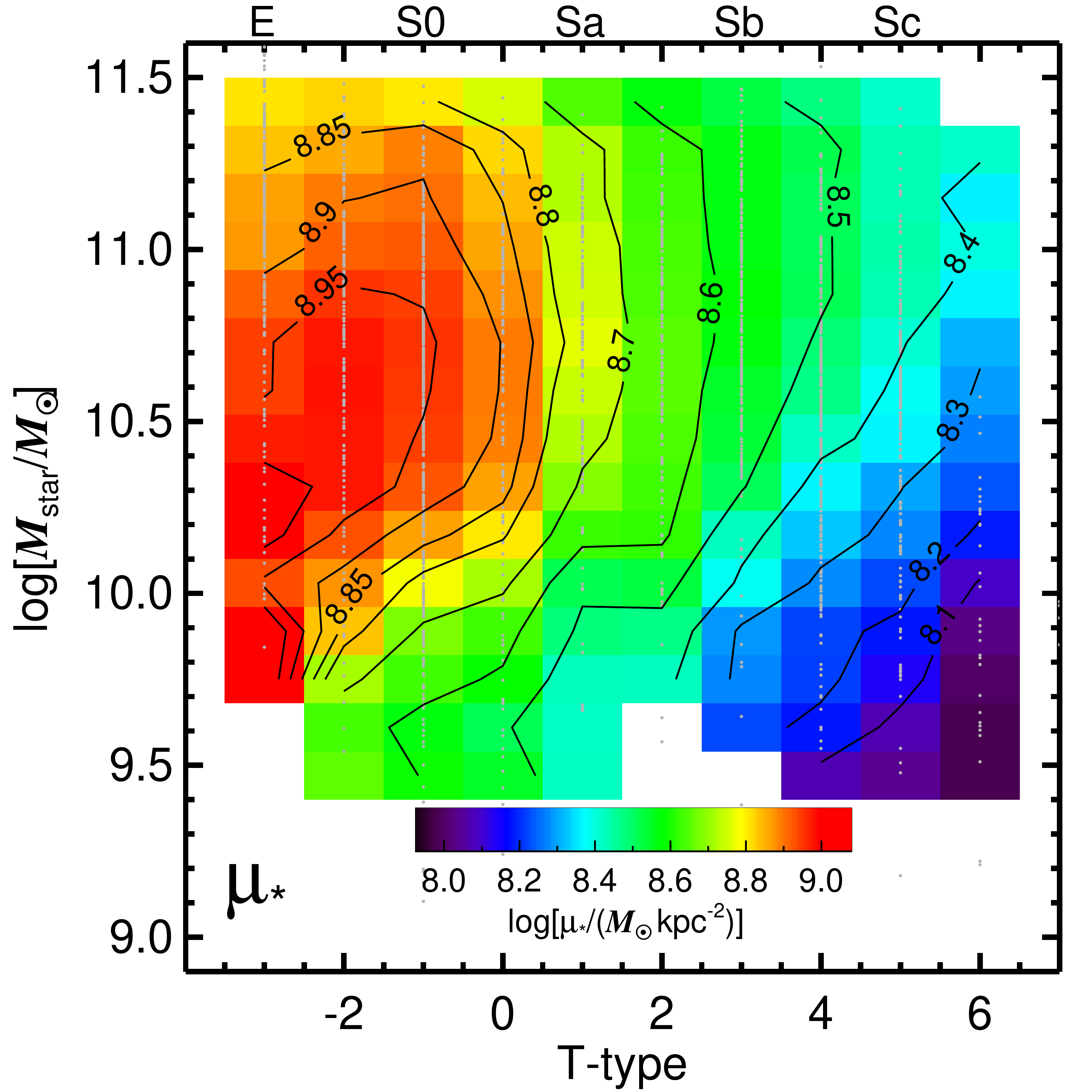}  
\centering
\caption{Central surface stellar mass density $\log\mu_*$ in the $T$-type versus $\log M_\mathrm{star}$ plane. Here, $\log\mu_*$ is represented by colors (see the color bars for the color-coded $\log\mu_*$) and contours (values in the middle of the contour lines indicate $\log\mu_*$). Other descriptions about the construction of the color map and contours are identical to those in Figure \ref{fig:slope}. The variation of $\log\mu_*$ is very similar to that of $s_\mathrm{in}$ shown in the right panel of Figure \ref{fig:slope}. The corresponding section: Section \ref{sec:discuss}.
\label{fig:mden}}
\end{figure} 

Previous studies suggested that baryon mass (or luminous matter) dominates the gravitational potentials of the inner regions of galaxies so that inner slopes of RCs are largely coupled with the central surface brightness or central surface stellar mass density ($\mu_*$) of galaxies \citep{Corradi1990,Noordermeer2007,Swaters2009,Lelli2013,Erroz2016}. Thus, we calculated $\mu_*$ for our galaxies to examine whether $\mu_*$ can explain the trends of $s_\mathrm{in}$ in the $T$-type versus $\log M_\mathrm{star}$ plane. Here, $\mu_*$ is half the stellar mass divided by the surface area of the half-light ellipse 
\begin{equation}
\mu_*=\frac{0.5M_\mathrm{star}}{\pi (R_e\sqrt{q_p})^2},
\label{eq:mus}
\end{equation}
where $q_p$ is the axis ratio of each galaxy estimated from the 2D S{\'e}rsic model in the NSA catalog. 

Figure \ref{fig:mden} shows $\log\mu_*$ in the $T$-type versus $\log M_\mathrm{star}$ plane. It is obvious that the variation of $\log\mu_*$ in the figure is very similar to that of $s_\mathrm{in}$ (the right panel of Figure \ref{fig:slope}) in the sense that (1) $\log\mu_*$ is mainly dependent on $T$-type (earlier-type galaxies have a larger $\mu_*$); (2) $\log\mu_*$ tends to slightly increase with $M_\mathrm{star}$ for $T$-type $\ge3$; (3) For $T$-type $\lesssim0$, low-mass and the most massive galaxies have smaller $\log\mu_*$ than galaxies with $\log(M_\mathrm{star}/M_\odot)\sim10.5$.\footnote{In the standard $\Lambda$ dark matter cosmological models, massive early-type galaxies with $\log(M_\mathrm{star}/M_\odot)\gtrsim11.2$ have experienced multiple (dry) mergers in their formation histories, which can puff up their sizes \citep{Oogi2013,Yoon2017} and make their $\mu_*$ relatively small compared to that of lower-mass galaxies with $\log(M_\mathrm{star}/M_\odot)\sim10.5$.} This indicates that $s_\mathrm{in}$ is coupled with $\mu_*$ in our galaxies in the sense that galaxies with a high $\mu_*$ have steep inner slopes in their RCs ($C_P=0.47$ for $\log\mu_*$ and $s_\mathrm{in}$), which implies that the gravitational potentials of the inner regions of galaxies are dominated by luminous matter in the whole range of $M_\mathrm{star}$ and $T$-type we probed.

In order to understand the dependence of $s_\mathrm{in}$ (and also other parameters: $s_\mathrm{out}$, $R_t$, and $V(R_t)$) on $\mu_*$ more elaborately, we used simple models of a rotating baryonic disk within a dark matter halo and examined how their RCs depend on the size and mass of the baryonic disk. For the rotating disk model, we adopted an exponential disk model with a thickness of 0.1. The RC of the baryonic disk model ($V_\mathrm{baryon}$)\footnote{Here, $V_\mathrm{baryon}$ can be assumed to be the RC of the stellar component in the disk. We did not consider the gas component separately.} was derived by a method in \citet{Noordermeer2008}. 

For the density profile of the dark matter halo model, we adopted the Navarro--Frenk--White (NFW) profile \citep{Navarro1997}: 
\begin{equation}
\rho(r)=\frac{\rho_\mathrm{s}}{(r/r_s)(1+r/r_s)^2},
\label{eq:nfw}
\end{equation}
where $r_s$ is the scale radius and $\rho_\mathrm{s}$ is the characteristic density related to the critical density of the universe. Then, the RC of the dark matter halo model \citep{Navarro1997,Lang2017} is 
\begin{equation}
V_\mathrm{DM}^2(r)=\frac{V_{200}^2}{x}\frac{\ln(1+cx)-(cx)/(1+cx)}{\ln(1+c)-c/(1+c)},
\label{eq:nfwrc}
\end{equation}
where $x=r/r_{200}$ and Parameter $c$ is the concentration parameter of the NFW profile. $c$ is related to $r_s$ through $r_s=r_{200}/c$. Here, $r_{200}$ is the radius within which the mean density is 200 times the critical density of the universe and $V_{200}$ is the circular speed at $r_{200}$, so that  $r_{200}=V_{200}/(10H(z))$. $V_{200}$ is connected with $M_{200}$ through $V_{200}^3=10GH(z)M_{200}$, where $G$ is the gravitational constant and $M_{200}$ is the halo mass within $r_{200}$. The Hubble parameter $H(z)$ \citep{Lang2017} is
\begin{equation}
H(z)=H_0\sqrt{\Omega_\mathrm{m}(1+z)^3+\Omega_\Lambda},
\label{eq:hp}
\end{equation}
in which we use the typical redshift of our sample ($z=0.04$). Since numerical simulations demonstrate that $c$ is correlated with halo mass in such a way that low-mass halos are more concentrated \citep{Bullock2001,Wechsler2002}, we used a relation between $c$ and $M_{200}$ in \citet{Duffy2008}, which is
\begin{equation}
c=5.74\bigg(\frac{M_{200}}{2\times10^{12}h^{-1}M_\odot}\bigg)^{-0.097}.
\label{eq:con}
\end{equation}
Therefore, $V_\mathrm{DM}(r)$ is determined by a single parameter $M_{200}$ in this simple dark matter halo model. The total circular velocity of the baryonic disk and the dark matter halo model \citep{Burkert2010,Burkert2016,Lang2017,Genzel2020} is calculated by
\begin{equation}
V_\mathrm{circ}^2(r)=V_\mathrm{baryon}^2(r)+V_\mathrm{DM}^2(r).
\label{eq:rccom}
\end{equation}

The velocity dispersion in the stellar components of the disk can generate an outward pressure gradient that reduces the centripetal force. Thus, we should take into account the effect of the pressure support in RCs. Assuming the velocity dispersion $\sigma_0$ is constant in the exponential disk model in hydrostatic equilibrium, the observed RC \citep{Burkert2010,Burkert2016,Lang2017,Genzel2020,Kretschmer2021} is described by
\begin{equation}
V_\mathrm{obs}^2(r)=V_\mathrm{circ}^2(r) - 2\sigma_0^2\bigg(\frac{r}{R_d}\bigg),
\label{eq:rcob}
\end{equation}
in which $R_d$ is the exponential scale length, which is related to $R_e$ by $R_e=1.68R_d$. Since $V_\mathrm{baryon}(r)$ is determined by $R_e$\footnote{In this model, $R_e$ is equivalent to the half-mass radius.} of the exponential profile and the total mass of the baryonic disk model ($M_\mathrm{baryon}$), $V_\mathrm{obs}(r)$ of the simple model is a function of the four parameters ($R_e$, $M_\mathrm{baryon}$, $\sigma_0$, and $M_{200}$).

\begin{figure}
\includegraphics[width=\linewidth]{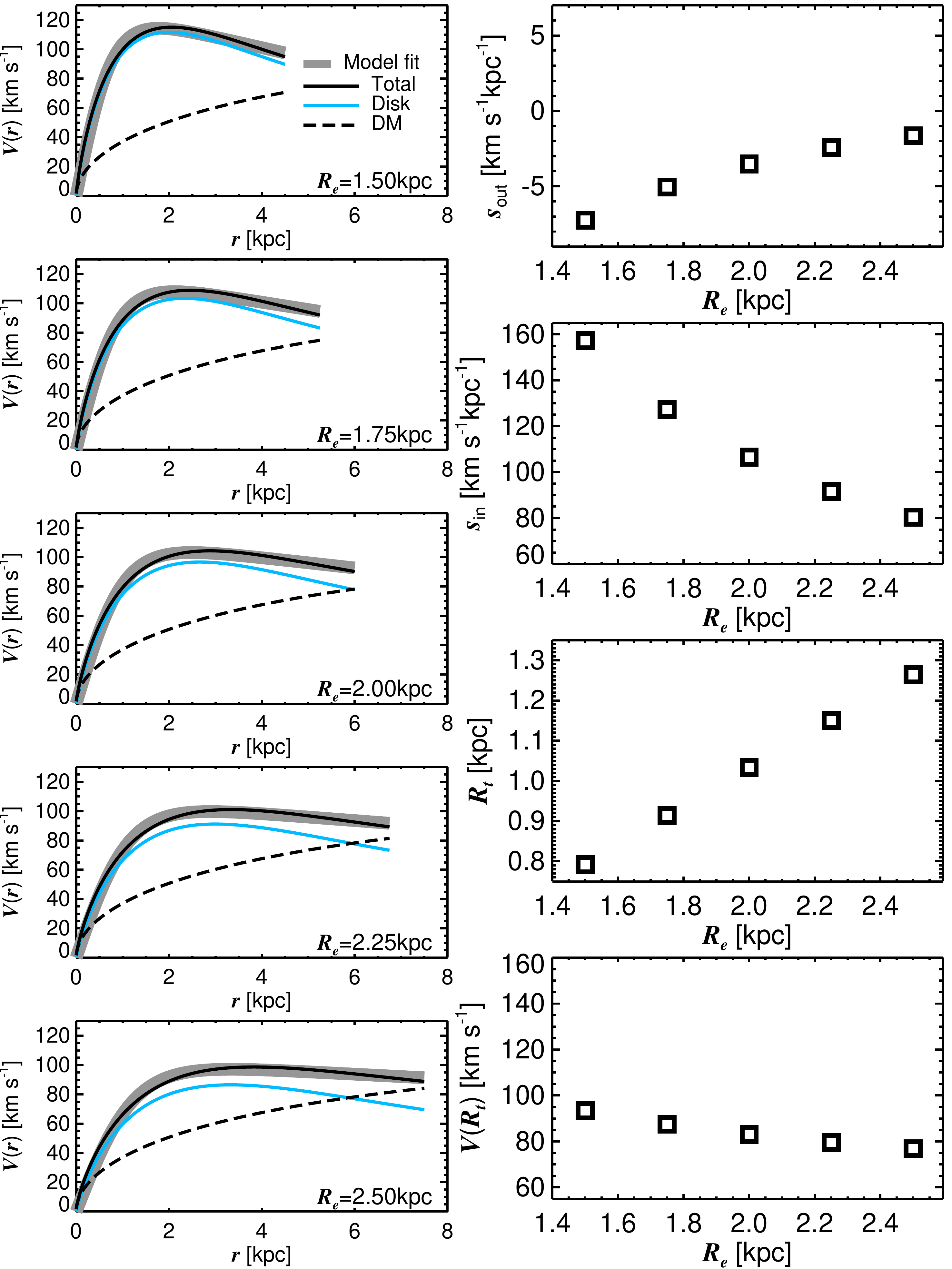}  %width=\linewidthscale=0.47
\centering
\caption{Dependence of the rotation curve shape and parameters ($s_\mathrm{out}$, $s_\mathrm{in}$, $R_t$, and $V(R_t)$) on $R_e$ of the baryonic rotating disk. Five values of $R_e$ (1.50, 1.75, 2.00, 2.25, and 2.50 kpc) are used here, while $M_\mathrm{baryon}$, $\sigma_0$, and $M_{200}$ are fixed to $7.5\times10^{9}M_\odot$, $20$ km s$^{-1}$, and $10^{11.5}M_\odot$, respectively. The black solid line means a total observed rotation curve of the model (Equation \ref{eq:rcob}). The contributions to the total rotation curve from the dark matter halo (Equation \ref{eq:nfwrc}) and the baryonic disk \citep{Noordermeer2008} are indicated by the dashed line and the blue solid line, respectively. The thick gray line is the fitted  functional form of Equation \ref{eq:rc1}. The rotation curve profiles are displayed up to $3R_e$. The corresponding section: Section \ref{sec:discuss}.
\label{fig:m_re}}
\end{figure}

\begin{figure}
\includegraphics[width=\linewidth]{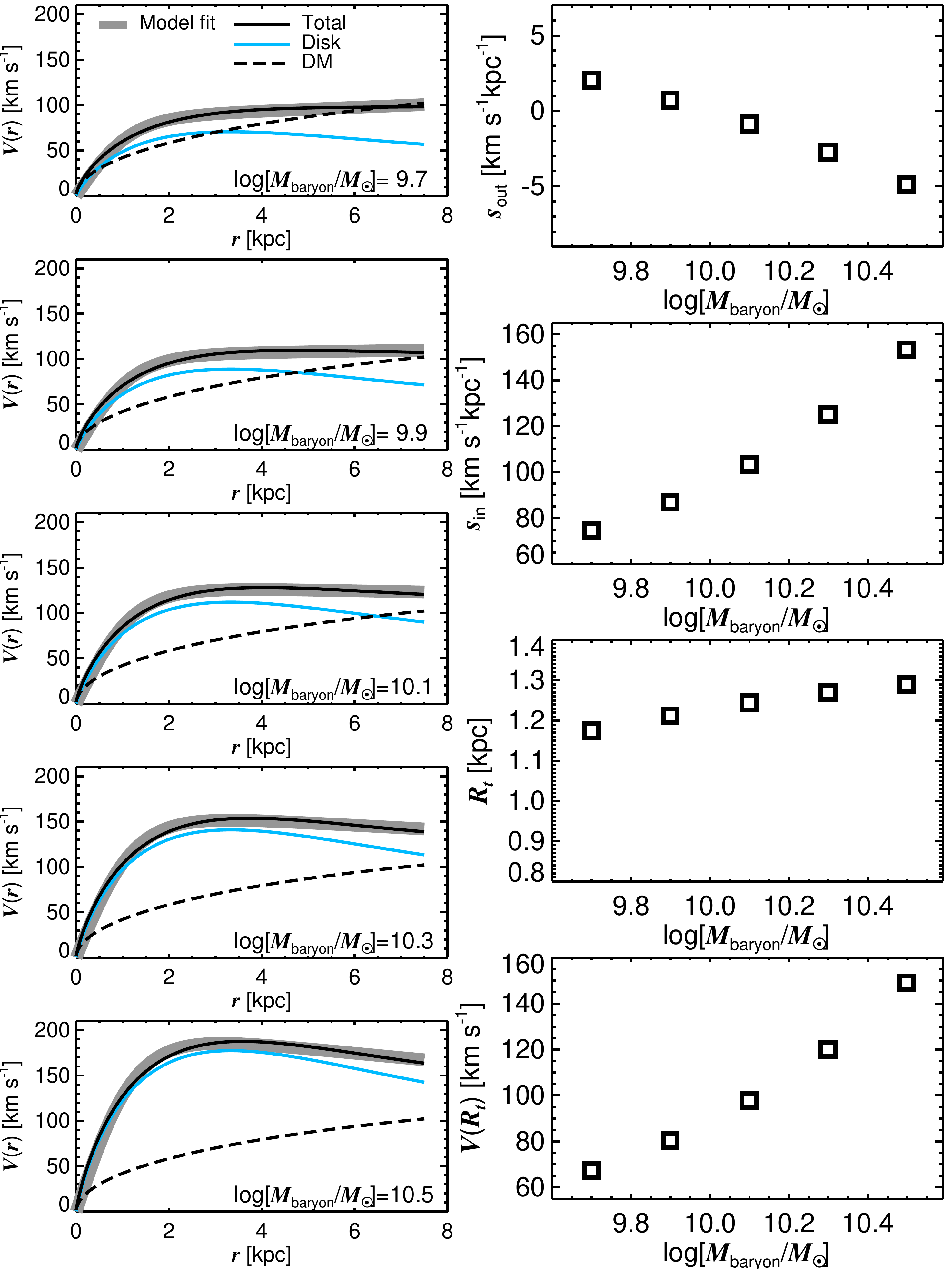}  %width=\linewidthscale=0.47
\centering
\caption{Dependence of the rotation curve shape and parameters ($s_\mathrm{out}$, $s_\mathrm{in}$, $R_t$, and $V(R_t)$) on $M_\mathrm{baryon}$ of the baryonic rotating disk. Five values of $M_\mathrm{baryon}$ ($10^{9.7}$, $10^{9.9}$, $10^{10.1}$, $10^{10.3}$, and $10^{10.5} M_\odot$) are used here, while $R_e$, $\sigma_0$, and $M_{200}$ are fixed to $2.5$ kpc, $20$ km s$^{-1}$, and $10^{12}M_\odot$, respectively. Other descriptions about the figure are identical to those in Figure \ref{fig:m_re}. The corresponding section: Section \ref{sec:discuss}.
\label{fig:m_bry}}
\end{figure} 

Figures \ref{fig:m_re} and \ref{fig:m_bry} show how the observed RC shape is dependent on $R_e$ (Figure \ref{fig:m_re}) or $M_\mathrm{baryon}$ (Figure \ref{fig:m_bry}) of the baryonic disk when the other three parameters are fixed in each case.\footnote{See the captions of Figures \ref{fig:m_re} and \ref{fig:m_bry} for parameters used in the models. We selected these parameter values to make model RCs broadly consistent with observational ones.} So, both of the figures also show a dependence of the RC shape on $\mu_*$ if we assume that $M_\mathrm{baryon}$ is similar to $M_\mathrm{star}$. The RC shape parameters such as $s_\mathrm{out}$, $s_\mathrm{in}$, $R_t$, and $V(R_t)$ were derived by fitting the functional form (Equation \ref{eq:rc1}) to the RCs of the simple models. The change of $R_e$ has a large effect on $R_t$ and $s_\mathrm{in}$ in such a way that $R_t$ increases and $s_\mathrm{in}$ decreases when $R_e$ goes up.\footnote{We note that the effect of the decrease in $R_e$ in the model is similar to that of the increase in the S{\'e}rsic index.} The variation of $M_\mathrm{baryon}$ largely affects $V(R_t)$ and $s_\mathrm{in}$ so that both $V(R_t)$ and $s_\mathrm{in}$ increase when $M_\mathrm{baryon}$ grows (see \citealt{Bekeraite2016} for the absolute magnitude--rotation velocity relation for rotating galaxies). As a result, our simple model verifies that $s_\mathrm{in}$ increases when $\mu_*$ increases ($R_e$ decreases and/or $M_\mathrm{baryon}$ increases), though it is naturally expected by the Newton's law. Moreover, the effects of $R_e$ and $M_\mathrm{baryon}$ on $R_t$ and $V(R_t)$ in the models are consistent with Figure \ref{fig:rv} and the explanations about it.

In Figures \ref{fig:m_re} and \ref{fig:m_bry}, the simple RC models also show that galaxies with a larger $\mu_*$ (smaller $R_e$ and/or larger $M_\mathrm{baryon}$) have smaller $s_\mathrm{out}$. This is due to the fact that concentrated matter in the central parts causes a more rapid decline of circular speed in the outer parts.\footnote{In the extreme case where most of the matter is concentrated in a very tiny region, the RC declines with distance just as the velocity of a planetary system does following the Kepler's laws  ($V(r)\propto r^{-0.5}$).} Furthermore, this is also attributable to the fact that galaxies with a larger $\mu_*$ are less affected by dark matter halos that make RCs rise in the outer regions unlike baryonic disks. We note that this is consistent with previous studies that suggested that outer RC profiles are related to light distributions or mass densities of galaxies \citep{Casertano1991,Kalinova2017,Tiley2019}. Our observational data also support the correlation between $\mu_*$ and $s_\mathrm{out}$ ($C_P=-0.37$ for $\log\mu_*$ and $s_\mathrm{out}$).

\begin{figure}
\includegraphics[width=\linewidth]{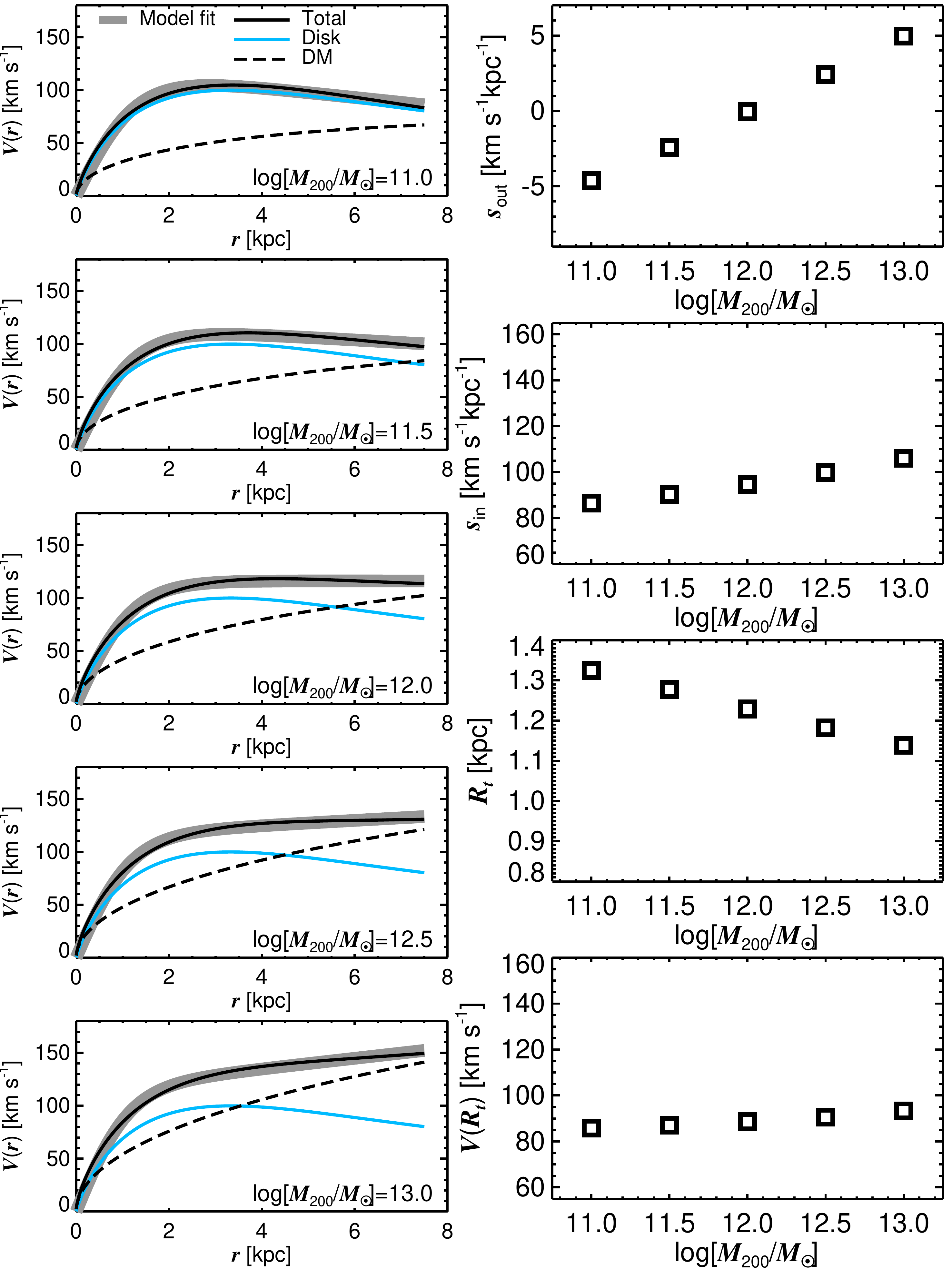}  %width=\linewidthscale=0.47
\centering
\caption{Dependence of the rotation curve shape and parameters ($s_\mathrm{out}$, $s_\mathrm{in}$, $R_t$, and $V(R_t)$) on $M_{200}$ of the dark matter halo. Five values of $M_{200}$ ($10^{11.0}$, $10^{11.5}$, $10^{12.0}$, $10^{12.5}$, and $10^{13.0} M_\odot$) are used here, while $R_e$, $M_\mathrm{baryon}$, and $\sigma_0$ are fixed to $2.5$ kpc, $10^{10}M_\odot$, and $20$ km s$^{-1}$, respectively. Other descriptions about the figure are identical to those in Figure \ref{fig:m_re}. The corresponding section: Section \ref{sec:discuss}.
\label{fig:m_dm}}
\end{figure} 

We also conducted a test to examine how the observed RC profiles depend on $M_{200}$ when the other three parameters are fixed, which is shown in Figure \ref{fig:m_dm}.\footnote{See the caption of Figure \ref{fig:m_dm} for parameters used in the test.} The test in Figure \ref{fig:m_dm} shows that the change of $M_{200}$ mainly has an effect on $s_\mathrm{out}$ in the sense that a higher fraction of dark matter makes a positively steeper slope in the outer RC profile. Therefore, the strong $M_\mathrm{star}$ dependence of $s_\mathrm{out}$ in late-type galaxies can be explained if lower-mass late-type galaxies are more dominated by dark matter than are higher-mass ones, which is supported by observational evidence from previous studies \citep{Casertano1991,Persic1996,Lelli2013,Martinsson2013,Courteau2015}. 

\begin{figure}
\includegraphics[width=\linewidth]{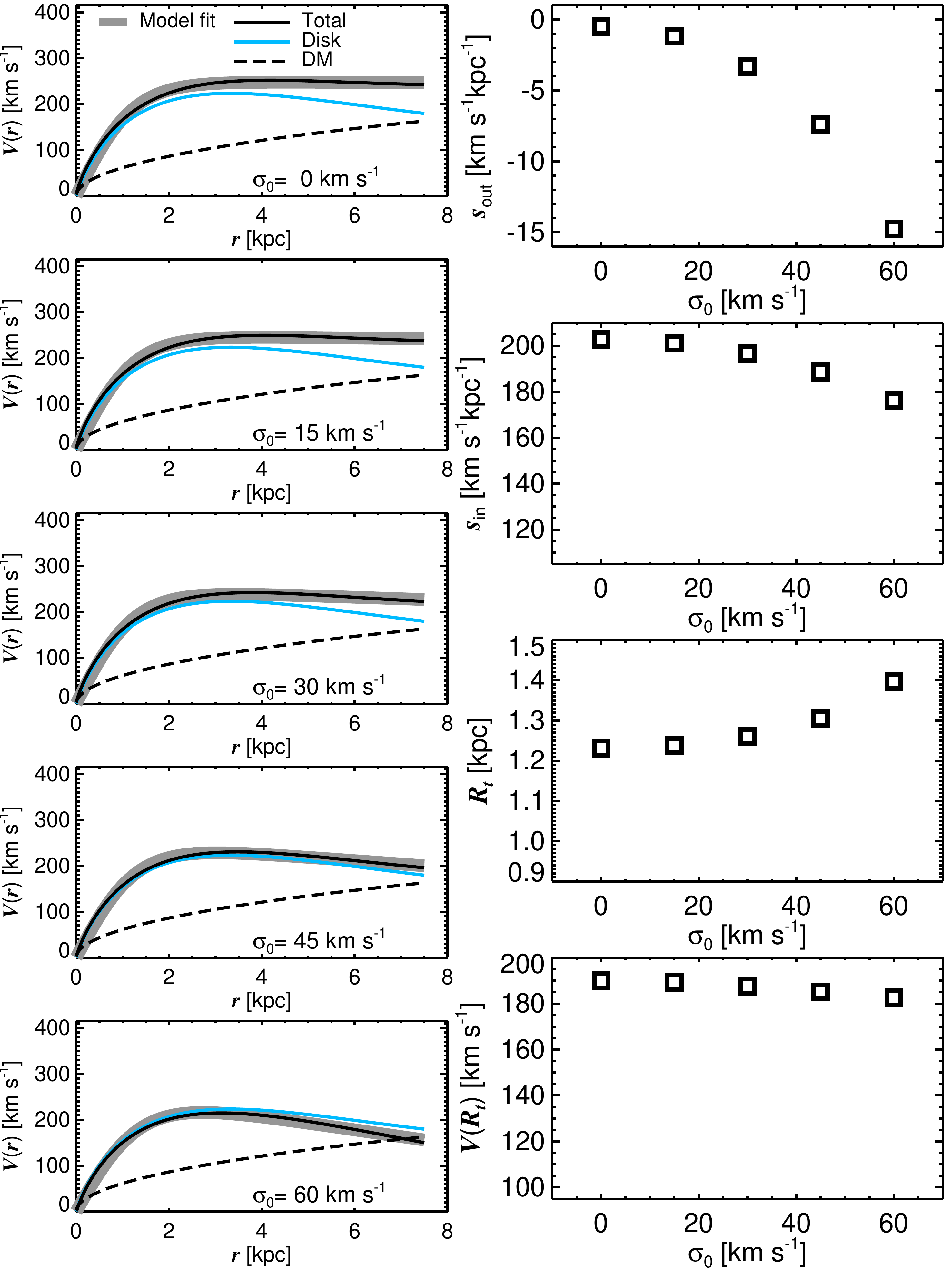}  %width=\linewidthscale=0.47
\centering
\caption{Dependence of the rotation curve shape and parameters ($s_\mathrm{out}$, $s_\mathrm{in}$, $R_t$, and $V(R_t)$) on $\sigma_0$ of the baryonic disk. Five values of $\sigma_0$ ($0$, $15$, $30$, $45$, and $60$ km s$^{-1}$) are used here, while $R_e$, $M_\mathrm{baryon}$, and $M_{200}$ are fixed to $2.5$ kpc, $5\times10^{10}M_\odot$ and $10^{13.5}M_\odot$, respectively. Other descriptions about the figure are identical to those in Figure \ref{fig:m_re}. The corresponding section: Section \ref{sec:discuss}.
\label{fig:m_vd}}
\end{figure} 

For a further test, we investigated how the observed RC profiles depend on $\sigma_0$ when the other three parameters of the model are fixed.\footnote{See the caption of Figure \ref{fig:m_vd} for parameters used in the test.} This test, the result of which is shown in Figure \ref{fig:m_vd}, demonstrates that the change in $\sigma_0$ largely affects $s_\mathrm{out}$ so that a model galaxy with a large $\sigma_0$ has a negatively steep slope in the outer RC profile. This can be the reason for our result that elliptical galaxies for which stellar velocity dispersions are high have descending RCs at large radii.

It is known that cluster environments can strip dark matter halos of satellite galaxies up to $80\%$ \citep{Whitmore1988,Whitmore1989,Smith2016,Rhee2017,Niemiec2019}. Thus, this environmental effect can contribute in part to the small values of $s_\mathrm{out}$ in elliptical galaxies of $T$-type $\le-2$, since they usually reside in dense environments according to the morphology--density relation \citep{Dressler1980,Springel2001,Thomas2006}.

Previous studies showed that some elliptical galaxies have irregular rotating features such as kinematically distinct cores and counter-rotating disks \citep{Cappellari2016,Graham2018}. Since their $s_\mathrm{out}$ is small compared with that of regularly rotating galaxies, the statistically small $s_\mathrm{out}$ in galaxies with $T$-type $\le-2$ can be partly due to the existence of such galaxies with irregular rotations.
\\

\section{Summary}\label{sec:summary}

We investigated stellar RCs of SDSS-IV MaNGA galaxies and their dependence on stellar mass and morphology ($T$-type). Deconvolution was conducted on the MaNGA IFU data using the LR algorithm in order to mitigate the seeing effect as in \citet{Chung2020}. Line-of-sight stellar velocities were extracted from the IFU data using the pPXF code that implements full spectrum fitting on galaxy spectra. To quantify shapes of RCs, we used a functional form of the RC (Equation \ref{eq:rc1}) that can model non-flat RCs at large radii and well represents RCs of MaNGA galaxies. Using this functional form, we fit 2D RC models (Equation \ref{eq:rc2}) to the line-of-sight velocity maps extracted using pPXF. We classified the morphologies of our galaxies and assigned $T$-type numbers. The number of galaxies in the final sample is 1940. The main conclusions about RCs of MaNGA galaxies are as follows.

\begin{enumerate}

\item Within the radial coverage of the MaNGA data, the flat RC at large radii applies only to massive late-type galaxies ($T$-type$\ge1$ and $\log(M_\mathrm{star}/M_\odot)\gtrsim10.8$) and S0 galaxies ($T$-type$=-1$ or $0$, and $\log(M_\mathrm{star}/M_\odot)\gtrsim10.0$). 

\item For late-type galaxies with $T$-type $\ge1$, galaxies with a lower $M_\mathrm{star}$ have a positively steeper outer slope ($s_\mathrm{out}$) so that $s_\mathrm{out}$ increases to about $+9$ km s$^{-1}$kpc$^{-1}$ at $\log(M_\mathrm{star}/M_\odot)\approx9.7$.  

\item By contrast, elliptical galaxies ($T$-type $\le-2$) have descending RCs at large radii. The $s_\mathrm{out}$ of elliptical galaxies falls down as $M_\mathrm{star}$ decreases so that $s_\mathrm{out}$ becomes as negative as $-15$ km s$^{-1}$kpc$^{-1}$ at $\log(M_\mathrm{star}/M_\odot)\approx10.2$. 

\item The inner slope of the RC ($s_\mathrm{in}$) is highest for elliptical galaxies with $T$-type $\lesssim-2$ and $\log(M_\mathrm{star}/M_\odot)\approx10.5$, and decreases as $T$-type increases or $M_\mathrm{star}$ changes away from $\log(M_\mathrm{star}/M_\odot)\approx10.5$.

\item The velocity at the turnover radius ($R_t$) is higher for a higher $M_\mathrm{star}$, and $R_t$ is larger for a higher $M_\mathrm{star}$ and later $T$-types.

\item Our observational data and the simple model of a rotating baryonic disk within a dark matter halo suggest that $s_\mathrm{in}$ is coupled with the central surface stellar mass density $\mu_*$: galaxies with a higher $\mu_*$ have a steeper $s_\mathrm{in}$, which implies that the gravitational potentials of the inner regions of galaxies are dominated by luminous matter (baryonic matter).

\end{enumerate}

We plan to conduct a further study using MaNGA data to investigate the effects of environments on RCs, since the environmental effect can also be an important factor for the shapes of RCs \citep{Whitmore1988,Whitmore1989,Smith2016,Rhee2017,Niemiec2019}.
\\

\begin{acknowledgments}
This work was supported by a KIAS Individual Grant PG076301 at the Korea Institute for Advanced Study.
Funding for the Sloan Digital Sky 
Survey IV has been provided by the 
Alfred P. Sloan Foundation, the U.S. 
Department of Energy Office of 
Science, and the Participating 
Institutions. 

SDSS-IV acknowledges support and 
resources from the Center for High 
Performance Computing  at the 
University of Utah. The SDSS 
website is www.sdss.org.

SDSS-IV is managed by the 
Astrophysical Research Consortium 
for the Participating Institutions 
of the SDSS Collaboration including 
the Brazilian Participation Group, 
the Carnegie Institution for Science, 
Carnegie Mellon University, Center for 
Astrophysics | Harvard \& 
Smithsonian, the Chilean Participation 
Group, the French Participation Group, 
Instituto de Astrof\'isica de 
Canarias, The Johns Hopkins 
University, Kavli Institute for the 
Physics and Mathematics of the 
Universe (IPMU) / University of 
Tokyo, the Korean Participation Group, 
Lawrence Berkeley National Laboratory, 
Leibniz Institut f\"ur Astrophysik 
Potsdam (AIP),  Max-Planck-Institut 
f\"ur Astronomie (MPIA Heidelberg), 
Max-Planck-Institut f\"ur 
Astrophysik (MPA Garching), 
Max-Planck-Institut f\"ur 
Extraterrestrische Physik (MPE), 
National Astronomical Observatories of 
China, New Mexico State University, 
New York University, University of 
Notre Dame, Observat\'ario 
Nacional / MCTI, The Ohio State 
University, Pennsylvania State 
University, Shanghai 
Astronomical Observatory, United 
Kingdom Participation Group, 
Universidad Nacional Aut\'onoma 
de M\'exico, University of Arizona, 
University of Colorado Boulder, 
University of Oxford, University of 
Portsmouth, University of Utah, 
University of Virginia, University 
of Washington, University of 
Wisconsin, Vanderbilt University, 
and Yale University.
\end{acknowledgments}

\end{document}